\documentclass[prd, superscriptaddress, tightenlines, longbibliography, nofootinbib, eqsecnum, amsfonts, amsmath, floatfix, notitlepage]{revtex4-2}
\usepackage[utf8]{inputenc}
\usepackage{mathrsfs}
\usepackage{euscript}
\usepackage{epsfig}
\usepackage{graphics}
\usepackage{graphicx}
\usepackage{amsmath}
\usepackage{amssymb}
\usepackage{timestamp}
\usepackage{bm}
\usepackage[usenames,dvipsnames,svgnames,table]{xcolor}
\usepackage{xspace}
\usepackage{wasysym}
\usepackage{times}
\usepackage{appendix}
\usepackage{lipsum}
\usepackage[nolist,nohyperlinks]{acronym}
\usepackage{float}
\usepackage{simplewick}
\usepackage{natbib, ifthen}
\usepackage{hyperref}

\newcommand{\la}{\langle}

\newcommand{\mksym}[1]{\ifmmode {\rm #1}\else #1\fi}

\newcommand{\lcdm}{{$\rm{\Lambda CDM}$}}

\newcommand{\mnu}{\sum m_\nu}

\providecommand{\lea}{\la}
\providecommand{\gea}{\ga}

\providecommand{\alt}{\lea}
\providecommand{\agt}{\gea}
\providecommand{\text}[1]{\rm{#1}}

\newcommand{\Mpc}{\text{Mpc}}

\newcommand{\grad}{\nabla}

\newcommand{\Hunit}{~\text{km}~\text{s}^{-1} \Mpc^{-1}}

\providecommand{\muK}{\mu{\rm K}}
\providecommand{\arcmin}{{\rm arcmin}}
\newcommand{\muKarcmin}{\,\muK\,\arcmin}

\newcommand{\eV}{\,\text{eV}}

\providecommand{\CAMB}{{\tt camb}}

\providecommand{\HALOFIT}{{\tt halofit}}

\newcommand{\begm}{\begin{pmatrix}}
\newcommand{\enm}{\end{pmatrix}}

\newcommand\ba{\begin{eqnarray}}
\newcommand\ea{\end{eqnarray}}
\newcommand\bea{\begin{eqnarray}}
\newcommand\eea{\end{eqnarray}}

\newcommand\be{\begin{equation}}
\newcommand\ee{\end{equation}}

\newcommand{\valpha}{{\boldsymbol{\alpha}}}
\newcommand{\vgrad}{{\boldsymbol{\nabla}}}

\newcommand{\vtheta}{\boldsymbol{\theta}}

\newcommand{\vell}{{\boldsymbol{\ell}}}

\newcommand{\boldvec}[1]{{\mbox{\boldmath{$#1$}}}}

\newcommand{\vL}{\boldvec{L}}

\newcommand{\vk}{\boldvec{k}}

\newcommand{\clo}{\mathcal{O}}

\definecolor{ZurichBlue}{rgb}{.255,.41,.884} 		% RoyalBlue of svgnames
\definecolor{ZurichRed}{rgb}{0.9, 0.1, 0} 			% Red of svgnames
\definecolor{ZurichGreen}{rgb}{.196,.504,.396} 		% LimeGreen of svgnames
\definecolor{ZurichYellow}{rgb}{1,.648,0} 			% Orange of svgnames
\definecolor{dodgerblue}{rgb}{0.12, 0.56, 1.0}
\definecolor{azure}{rgb}{0.0, 0.5, 1.0}
\definecolor{awesome}{rgb}{1.0, 0.13, 0.32}
\definecolor{alizarincrimson}{rgb}{0.82, 0.1, 0.26}
\definecolor{mediumpurple}{rgb}{0.58, 0.44, 0.86}
\definecolor{lasallegreen}{rgb}{0.03, 0.47, 0.19}

\DeclareMathAlphabet{\pazocal}{OMS}{zplm}{m}{n}
\newcommand{\Mkappa}{\pazocal{M}}
\newcommand{\Fkappa}{\pazocal{F}}

\begin{document}

%%%%%%%%%%%%%%%%%%%%%%%%%%%%%%%%%%%%%%% Some newcommands %%%%%%%%%%%%%%%%%%%%%%%%%%%%%%%%%%%%
\newcommand{\ie}{i.e.}
\newcommand{\etal}{\textit{et al.}}
\newcommand{\rmi}{{\rm i}}

\newcommand{\ellp}{\ell^{\prime}}
\newcommand{\bfell}{{\pmb{\ell}}}
\newcommand{\bfellp}{{\pmb{\ell}^{\prime}}}
\newcommand{\bfellpp}{{\pmb{\ell}^{\prime \prime}}}

\newcommand{\bfA}{{\bf{A}}}
\newcommand{\bfe}{{\bf{e}}}
\newcommand{\bfr}{{\bf{r}}}
\newcommand{\bfn}{{\bf{n}}}
\newcommand{\bfnp}{{\bf{n}}^{\prime}}
\newcommand{\bfx}{{\bf{x}}}
\newcommand{\bfk}{{\bf{k}}}
\newcommand{\bfd}{{\bf{d}}}
\newcommand{\bfC}{{\bf{C}}}
\newcommand{\bfL}{{\bf{L}}}
\newcommand{\bfLp}{{\bf{L}}^{\prime}}
\newcommand{\bfLpp}{{\bf{L}}^{\prime \prime}}
\newcommand{\bfLppp}{{\bf{L}}^{\prime \prime \prime}}

\newcommand{\bfmu}{{\bm\mu}}
\newcommand{\bftgr}{{\bm\theta}^{\textrm{GR}}}
\newcommand{\bftmg}{{\bm\theta}^{\textrm{MG}}}
\newcommand{\bft}{{\bm\theta}}
\newcommand{\bfa}{{\bm\alpha}}
\newcommand{\bftp}{{\bm\theta}^{\prime}}

\newcommand{\calA}{{\mathcal{A}}}
\newcommand{\calC}{{\mathcal{C}}}
\newcommand{\calD}{{\mathcal{D}}}
\newcommand{\calR}{{\mathcal{R}}}
\newcommand{\calO}{{\mathcal{O}}}
\newcommand{\calS}{{\mathcal{S}}}

\newcommand{\dlw}{{(2 \pi)^2}}

\newcommand{\lef}{{g (\chi , \chi^{\prime})}}
\newcommand{\chip}{{\chi^{\prime}}}
\newcommand{\chipp}{{\chi^{\prime \prime}}}
\newcommand{\chippp}{{\chi^{\prime \prime \prime}}}

\renewcommand{\Re}{\operatorname{Re}}
\renewcommand{\Im}{\operatorname{Im}}

\newcommand\GReq{\mathrel{\overset{\makebox[0pt]{\mbox{\normalfont\tiny\sffamily GR}}}{=}}}

\def\n{\noindent}

% Uses Tikz package
%\def\checkmark{\tikz\fill[scale=0.4](0,.35) -- (.25,0) -- (1,.7) -- (.25,.15) -- cycle;}

\newcommand{\Sussex}{Department of Physics \& Astronomy, University of Sussex, Brighton BN1 9QH, UK}

%%%%%%%%%%%%%%%%%%%%%%%%%%%%%%%%%%%%%%%%%%%%%%%%%%%%%%%%%%%%%%%%%%%%%%%%%%%%

\title{Impact of post-Born lensing on the CMB}

\author{Geraint Pratten}
\affiliation{\Sussex}
\author{Antony Lewis}
\affiliation{\Sussex}
\homepage{http://cosmologist.info}

%%%%%%%%%%%%%%%%%%%%%%%%%%Abstract%%%%%%%%%%%%%%%%%%%%%%%%%%%%%%%%%%%%%%%%%%%%%%%%%
\timestamp
\begin{abstract}

Lensing of the CMB is affected by post-Born lensing, producing corrections to the convergence power spectrum and
introducing field rotation. We show numerically that the lensing convergence power spectrum is affected at the $\alt 0.2\%$ level on accessible scales, and that this correction and the field rotation are negligible
for observations with arcminute beam and noise levels $\agt 1 \muKarcmin $. The field rotation generates $\sim 2.5\%$ of the total lensing B-mode polarization amplitude ($0.2\%$ in power on small scales), but has a blue spectrum on large scales, making it highly subdominant to the convergence B modes on scales where they are a source of confusion for the signal from primordial gravitational waves.
Since the post-Born signal is non-linear, it also generates a bispectrum with the convergence. We show that the post-Born contributions to the bispectrum substantially change the shape predicted from large-scale structure non-linearities alone, and hence must be included to estimate the expected total signal and impact of bispectrum biases on CMB lensing reconstruction quadratic estimators and other observables.
The field-rotation power spectrum only becomes potentially detectable for noise levels $\ll 1 \muKarcmin$, but its bispectrum with the convergence may be observable at $\sim 3\sigma$ with Stage IV observations.
Rotation-induced and convergence-induced B modes are slightly correlated by the bispectrum, and the bispectrum also produces additional contributions to the lensed BB power spectrum.
\end{abstract}

%%%%%%%%%%%%%%%%%%%%%%%%%%%%%%%%%%%%%%%%%%%%%%%%%e%%%%%%%%%%%%%%%%%%%
\pacs{
}

\maketitle
%%%%%%%%%%%%%%%%%%%%%%%%%Introduction%%%%%%%%%%%%%%%%%%%%%%%%%%%%%%%%%%%%%%%%%%%%%%%

\begin{acronym}
\acrodef{WL}[WL]{Weak Lensing}
\end{acronym}

\newcommand{\WL}{\ac{WL}\xspace}

\section{Introduction}
\label{sec:intro}
Weak lensing of the cosmic microwave background (CMB) has now been detected at high significance, and is starting to be a valuable tool improve cosmological parameter constraints~\cite{Ade:2015zua}. Since the CMB lensing kernel is broadly peaked at high redshift, the lensing potential is nearly linear and Gaussian, with only modest dependence on non-linear structure growth. Perturbative models of the non-linear matter power spectrum, combined with a small uncertainty from a tail of strongly non-linear contributions, lead to lensing power spectrum predictions that are under control and accurate enough for the coming generation of experiments~\cite{Baldauf:2016sjb}. However a number of other effects are potentially important, including biases in lensing quadratic estimators from a variety of correlated and non-Gaussian sources~\cite{vanEngelen:2013rla,Osborne:2013nna,Bohm:2016gzt}, which must also be carefully accounted for.

In this paper we calculate post-Born corrections to CMB lensing, which describe the non-linear effect of interaction between more than one lensing deflection along the line of sight. This can lead to a quantitative change to the lensing potential power spectrum that could be important on small scales.
Recently Ref.~\cite{Hagstotz:2014qea} has claimed there are large effects on the CMB lensing power, even though earlier work has shown the effects are expected to be small and do not grow dramatically on small scales~\cite{Cooray:2002mj,Krause:2009yr}. Since there is no fully correct numerical calculation in the literature for CMB lensing, we will revisit the calculation and explain the expected scale-dependence of the signal and hence why the effect is indeed small.

Although the effect on the convergence spectrum turns out to be negligible for near-future observations, there are other potentially interesting effects. In particular, the composition of two unaligned lensing shears leads to a qualitatively different anti-symmetric field rotation component that is not present at linear order. It has been argued that the field rotation may ultimately limit the precision with which delensing can be used to extract small primordial gravitational wave signal from CMB B-mode polarization, though the effect is subdominant for currently planned experiments~\cite{Hirata:2003ka} and can also be delensed.

Since the post-Born effects are non-linear, they will also introduce a non-Gaussian signal.
Non-Gaussianity is usually expected to be dominated by the bispectrum of non-linear large-scale structure (LSS), though this is substantially suppressed in the lensing signal due to the large number of lenses along the line of sight to the CMB. The contribution from post-Born effects to the convergence bispectrum is known to be small but potentially important for galaxy lensing~\cite{Dodelson:2005rf}; for CMB lensing the contribution could be relatively much more important since the potentials are more linear and the path length to last scattering is longer.
The bispectrum signal is interesting in its own right, and potentially an additional source of cosmological information~\cite{Namikawa:2016jff}, as well as a possible source of bias in CMB lensing power spectrum estimators~\cite{Bohm:2016gzt}. We show that the post-Born contribution to the CMB convergence bispectrum is of comparable size to the LSS bispectrum, and substantially changes the shape of the total bispectrum signal.
Post-Born effects also generate a new mixed bispectrum between the convergence and field rotation, which we show may become detectable with future observations.

The outline of this paper is as follows. In Sec.~\ref{sec:intro} we start by establishing notation and reviewing standard results, then describe the corrections to the lensing distortion tensor from post-Born corrections up to third order in the gravitational potential. In Sec.~\ref{sec:lenspower} we calculate the corrections to the CMB lensing convergence power spectrum and the field rotation power spectrum, and discuss their observational relevance.
In Sec.~\ref{sec:PBBispectrum} we discuss the post-Born bispectrum, how it modifies the bispectrum from large-scale structure growth, a new distinctive mixed convergence-rotation bispectrum, and the detectability and observational importance of these bispectra.
In Sec.~\ref{sec:CMBpower} we asses the impact on the lensed CMB power spectra, including new contributions proportional to the mixed bispectrum.
We finish with conclusions, and relegate some details of the post-Born calculations to appendices. Throughout we assume a flat statistically-isotropic and homogenous \lcdm\ universe evolving according to General Relativity and use natural units ($c=1$). Example numerical results are based on linear power spectra from \CAMB~\cite{Lewis:1999bs}, with baryon density $\Omega_b h^2=0.022$, dark matter density $\Omega_c h^2=0.122$, scalar perturbation power at $0.05\Mpc^{-1}$ of $A_s=2\times 10^{-9}$ with constant spectral index $n_s=0.965$, Hubble parameter $H_0 =67 \Hunit$ and one minimal massive neutrino with $\mnu=0.06\eV$. Corrections to the matter power spectrum from non-linear growth are modelled by using \HALOFIT~\cite{Smith:2002dz,Takahashi:2012em}. Since post-Born effects are small on large scales we use the flat-sky and Limber approximations throughout.

\section{CMB lensing potentials}

\subsection{The Lens Equation}
\label{sec:intro}
{
In weak gravitational lensing, a photon emitted by some distant source will follow geodesics that are perturbed with respect to the background due to intervening gravitational potentials. The aim is therefore to find some expression for the deflection angle in terms of the underlying  perturbations. Working in the conformal Newtonian gauge, a linearly perturbed FLRW spacetime can be written as
\begin{align}
ds^2 &= a^2 (\eta) \left[ - \left( 1 + 2 \Psi_N \right) d \eta^2 + \left( 1 + 2 \Phi_N \right) \gamma_{ij} dx^i dx^j \right] ,
\end{align}
\n
where $a$ is the scale factor, $\eta$ the conformal time, $\Psi_N$ and $\Phi_N$ are the scalar metric potentials and $\gamma_{ij}$ the unperturbed spatial metric
\begin{align}
\gamma_{ij} dx^i dx^j &= d \chi^2 + \chi^2 \left( d \theta^2 + \sin^2 \theta \, d \phi^2 \right) ,
\end{align}
\n
where $\chi$ is the comoving radial distance. As we are interested in calculating the paths of null geodesics, we will only be interested in the Weyl potential
\begin{align}
\Psi &= \frac{1}{2} \left( \Psi_N - \Phi_N \right) .
\end{align}
\n
At late times and in the absence of anisotropic stress, such as in the matter and dark energy dominated era, $\Phi_N = - \Psi_N = \Psi$. The Weyl potential can then be directly related to the comoving matter perturbations via the Poisson equation.

Weak gravitational lensing can be described by considering the difference in the paths taken by null geodesics in the background and our perturbed spacetime, $\bar{x}^{\mu}$ and $x^{\mu}$. Without loss of generality, we consider an observer at $\bfx = 0$ such that photons in the background spacetime will follow null geodesics parameterized by $\bar{x}^{\mu} = (\eta_0 - \chi,\chi \bft_{\mathcal{S}})$. The deflection angle $\alpha^a$ can be defined by projecting the angular components of the deviation vector $\delta \bfx = \bfx - \bar{\bfx}$ onto the sphere \cite{Yamauchi:2012bc}
\begin{align}
\alpha^a = \frac{\delta x^i \, e_i^{\phantom{i}a}}{\chi} ,
\end{align}
\n
where $e_i^{\phantom{i} a}$ are basis vectors orthogonal to $\bft$ and $a$ denotes angular coordinates $\lbrace \theta,\phi \rbrace$. As the deflection angle is just a vector, it can be decomposed into two irreducible components corresponding to two degrees of freedom
\begin{align}
\alpha_a &= \nabla_a \phi + \epsilon_{ab} \nabla^b \Omega ,
\label{eq:alphadecomp}
\end{align}
\n
where $\phi$ is the lensing potential, $\Omega$ the curl potential and $\boldsymbol{\nabla}_{} = \chi \boldsymbol{\nabla}_{\perp}$ is the covariant derivative on the sphere defined by $\bft$. In order to relate the deflection angle to the metric potentials, we have to obtain the deviation vector from the integral solutions to the null geodesic equation in the perturbed spacetime.

The angle $\bft_{}$ of an observed source is related to the unlensed source-plane angle $\boldsymbol{\theta}_{\mathcal{S}}$ by the deflection angle:
\begin{align}
{\theta}_{\mathcal{S}}^a (\bft , \chi) &= \theta^a_{} + \alpha^a .
\end{align}
\n
The linearized mapping between the source plane and the image planes is described by the Jacobian matrix $\mathcal{A}_{ab}$, which is defined by
\begin{align}
\label{eqn:Jacobian}
\mathcal{A}^{ab} &= \frac{\partial \theta^a_{\mathcal{S}}}{\partial \theta^b_{}} =
\begin{pmatrix}
 1 - \kappa - \gamma_1 & -\gamma_2 - \omega \\
  -\gamma_2 + \omega & 1 - \kappa + \gamma_1
 \end{pmatrix} .
\end{align}
\n
Here $\kappa$ is the lensing convergence, $\gamma = \gamma_1 + i \gamma_2$ is the complex lensing shear and $\omega$ is the image rotation, which vanishes at linear order. The weak lensing convergence and rotation can then be determined from the deflection angle $\boldsymbol{\alpha}$ by
\begin{align}
\kappa = - \frac{1}{2} \nabla^a {\alpha}_a = -\frac{1}{2} \vgrad^2 \phi, \quad {\rm{and}} \quad \omega = \frac{1}{2} \epsilon^{ab} \nabla_a {\alpha}_b = - \frac{1}{2} \vgrad^2 \Omega .
\end{align}
\n
As the deflection angle $\bm{\alpha}$ only has two degrees of freedom there is an equivalence between the convergence $\kappa$ and E-mode (gradient) shear $\gamma_E$, as well as between the curl $\omega$ and B-mode (curl) shear $\gamma_B$.

The deformation of the Jacobian matrix, denoted $\psi_{ab}$, encodes all the information regarding lensing and is defined by\footnote{Here the deformation tensor $\psi_{ab}$ is not to be confused with derivatives of the Weyl potential $\Psi$.} \cite{Bartelmann:1999yn}
\begin{align}
\label{eqn:deftensor}
\psi_{ab} &= \delta_{ab} - \mathcal{A}_{ab} .
\end{align}
\n
The convergence, shear and rotation can be expressed in terms of the distortion tensor $\psi_{ij}$ in the conventional way $\kappa = \frac{1}{2}(\psi_{11} + \psi_{22})$, $\gamma_1 = \frac{1}{2}(\psi_{11} - \psi_{22})$, $\gamma_2 = \frac{1}{2}(\psi_{12} + \psi_{21})$ and $\omega = \frac{1}{2}(\psi_{12} - \psi_{21})$.
}

\begin{figure}
\begin{center}
\includegraphics[width=140mm]{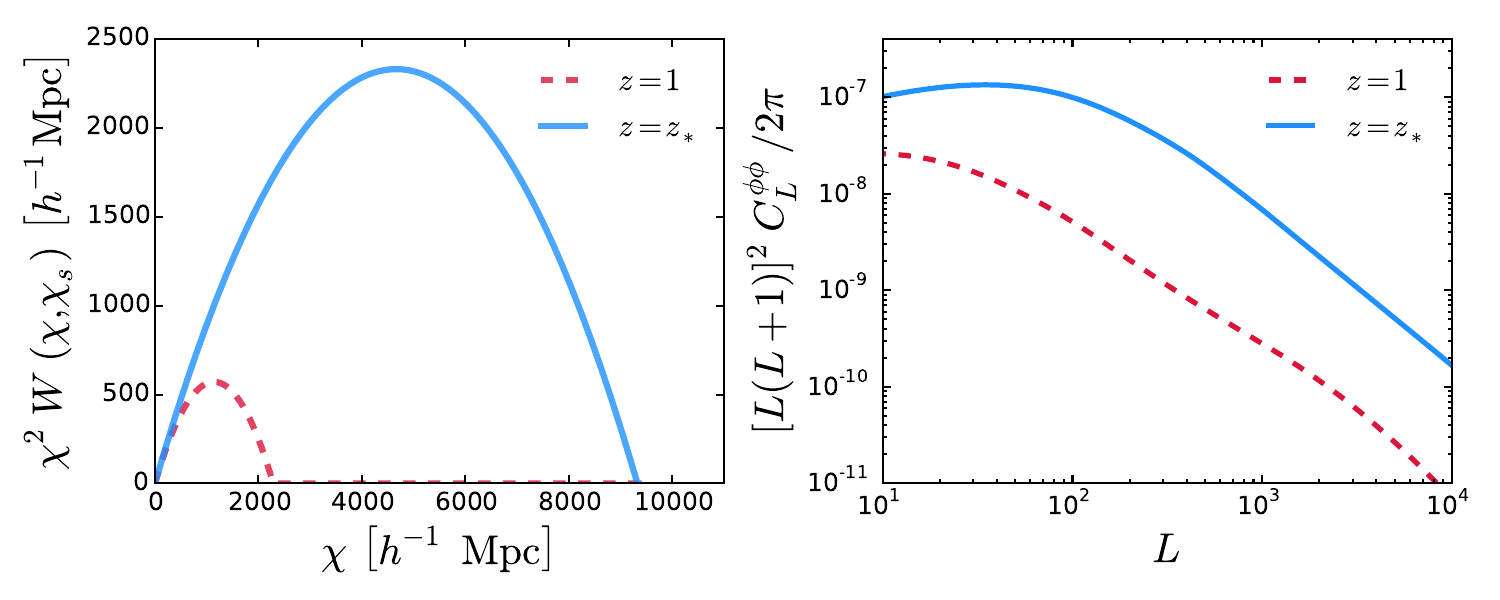}
\caption{
The left plots shows the lensing efficiency function $\chi^2 W(\chi,\chi_s)$ for $\chi_s$ at redshift $z=1$ (red, dashed) and $z = z_{\ast}$ (blue). The right plot shows the lensing potential power spectrum $C^{\phi\phi}_L$ generated using the non-linear Weyl potential power spectrum $P_{\Psi\Psi}$ for $z=1$ (red, dashed) and CMB lensing ($z=z_{\ast}$, blue). We plot $[L(L+ 1)]^2 C^{\phi \phi}_{L}/(2\pi)$, corresponding to the deflection angle variance per $\log L$, which is proportional to $C_L^{\kappa\kappa}$.
}
\label{fig:LEF}
\end{center}
\end{figure}

\subsection{Leading-order result}
{
The leading order contribution to lensing can be derived by adopting the Born approximation. In this approximation we implicitly assume that the photons travel along unperturbed geodesics $\bfx^{(0)} \approx \bft \chi$ and that the initial deformations, including any lens-lens coupling between lenses at different redshifts, are sufficiently small that we may assume $\psi_{ab}^{(0)} \approx 0$. Solving the null geodesic equation to linear order in the metric potentials, the deviation vector will be given by \cite{Lewis:2006fu}
\begin{align}
\delta x_a (\bft , \chi_s) &= -2 \int^{\chi_s}_0 d \chip \, \left( \chi_s - \chip \right) \, \Psi_{,a} (\bft, \chip) ,
\end{align}
\n
where commas denote spatial transverse derivatives $\boldsymbol{\nabla}_{\perp}$. At linear order, $\theta_{\mathcal{S}}^a$ is given by
\begin{align}
\bft_{\mathcal{S}} (\bft, \chi_s) &= \bft_{} - 2 \int^{\chi_s}_0 d\chi' \frac{( \chi_s - \chi^{\prime} )}{\chip \chi_s} \, \boldsymbol{\nabla}_{} \, \Psi{} (\bft , \chip) ,
\end{align}
\n
and the lensing deflection angle is therefore
\begin{align}
\boldsymbol{\alpha} &= -2 \int^{\chi_s}_0 d\chi' \frac{( \chi_s - \chi^{\prime} )}{\chip \, \chi_s} \, \boldsymbol{\nabla}_{} \Psi (\bft , \chip) .
\end{align}
\n
At linear order, the curl potential is identically zero and the deflection angle is just $\boldsymbol{\alpha} = \boldsymbol{\nabla}_{} \phi$, where the projected lensing $\phi$ potential is defined to be \cite{Lewis:2006fu}
\begin{align}
\phi &= -2 \int^{\chi_s}_0 d\chi' \frac{( \chi_s - \chi^{\prime} )}{\chip \, \chi_s} \, \Psi (\bft , \chip) ,
\end{align}
\n
Using Eqns.~(\ref{eqn:deftensor}) and ~(\ref{eqn:Jacobian}) together with the Born approximation leads to the first-order deformation tensor
\begin{align}
\psi_{ab}^{(1)} (\bft , \chi_s) &= 2 \int^{\chi_s}_0 d \chip \, \chi^{\prime 2} \, W( \chip, \chi_s) \, \Psi_{,ab} (\bft , \chip) ,
\end{align}
\n
where we have introduced the lensing efficiency function $W(\chip,\chi)$
\begin{align}
W(\chip,\chi) &= \left( \frac{1}{\chip} - \frac{1}{\chi} \right) \, \Theta (\chi - \chip) ,
\end{align}
\n
in which $\Theta (x)$ is the Heaviside step function. The concomitant lensing observables $\lbrace \kappa , \gamma_E , \gamma_B , \omega \rbrace$ may then be extracted by the appropriate contractions of the deformation tensor $\psi_{ab}$ along with the usual definition of the angular power spectrum between two fields $\Gamma$ and $\Gamma^{\prime}$, $\langle {\Gamma} (\bfell) {\Gamma}^{\prime} (\bfellp) \rangle = (2 \pi)^2 C_{\Gamma \Gamma^{\prime}} \delta_D (\bfell + \bfellp)$, see Appendix \ref{appendix:power} for further details.
}

In the flat-sky approximation, the 2D angular Fourier-transform at a fixed comoving radial distance $\chi$ is given by
\begin{align}
\Psi (\bft , \chi) &= \int \frac{d^2 \bfL}{(2 \pi)^2} \, {\Psi} (\bfL ; \chi) \, e^{i \bfL \cdot \bft}.
\end{align}
Consequentially, the spatial derivatives of the potential can be expressed in terms of the angular Fourier transform ${\Psi}$ \cite{Krause:2009yr}
\begin{align}
\Psi_{,ab \dots m} (\bft ; \chi) &= \frac{i^m}{\chi^m} \int \frac{d^2 \bfL}{( 2 \pi )^2} L_a L_b \dots L_m \, {\Psi} (\bfL ; \chi) \, e^{i \bfL \cdot \bft}.
\end{align}
Under the lowest-order Limber approximation the power spectrum of the Weyl potential is given by \cite{Kaiser92}
\begin{align}
\label{eqn:LimberPS}
\left\langle {\Psi} \left( \bfL ; \chi \right) \, {\Psi} \left( \bfL^{\prime} ; \chi^{\prime} \right) \right\rangle &= \left( 2 \pi \right)^2 \, \delta \left( \bfL + \bfL^{\prime} \right) \, \frac{\delta_D \left( \chi - \chi^{\prime} \right)}{\chi^2} \, P_{\Psi \Psi} \left( \frac{L}{\chi} , z(\chi) \right) ,
\end{align}
\n
where $P_{\Psi \Psi} (L / \chi , z(\chi))$ is the 3-dimensional Weyl potential power spectrum at a redshift of $z (\chi)$.

Combining these results, the linear lensing potential power spectrum for a source planes at $\chi_s$ and $\chi_s'$ (with $\chi_s'<\chi_s$) can be written as
\footnote{The lowest-order Limber approximation used here is not accurate at low $L$ for the linear result, and can easily be improved using the next lowest-order approximation~\cite{LoVerde:2008re}. However, for consistency here we used the lowest order result everywhere; the small post-Born corrections we calculate, which are only significant on small scales,  can be added to a an accurate fully-sky calculation of the linear result.}
\begin{align}
C^{\kappa\kappa}_L(\chi_s,\chi_s')= L^4\int^{\chi_s}_0 d \chi \, \frac{W (\chi , \chi_s) W (\chi , \chi_s')}{\chi^2} \, P_{\Psi \Psi} \left( \frac{L}{\chi} , z(\chi) \right),
\end{align}
where we can suppress the second argument when both source planes are the same.
For CMB lensing the last-scattering surface is well approximated by a single source plane at $z_s=z_*$, where $z_*$ is the redshift of the peak of the visibility,
and the CMB lensing potential power spectrum is given by $C_L^{\phi\phi} = (4/L^4)C_L^{\kappa\kappa}(z_*)$.
\n
At linear order, the curl potential $C^{\omega\omega}_{L}$ is identically zero.

Most work on post-Born corrections in the literature has focussed on galaxy lensing at $z\sim 1$ \cite{Cooray:2002mj,Dodelson:2005zj,Dodelson:2005rf,Shapiro:2006em,Hilbert:2008kb,Krause:2009yr,Thomas:2014aga}. Compared to galaxy lensing, the CMB lensing kernel peaks at substantially higher redshift, and the lensing potential power is larger due the greater path length to last scattering (see Fig.~\ref{fig:LEF}). The CMB lensing spectrum is sensitive to non-linear corrections to the matter (and hence Weyl) power spectrum, but at a much more modest level than galaxy lensing due to the greater dependence on higher redshifts. This and the increased lensing amplitude make the post-Born corrections potentially relatively more important for CMB lensing.

\subsection{Post-Born corrections}

Post-Born corrections account for the fact that photons do not travel along the same path as background geodesics, giving non-linear corrections to the distortion tensor. There are several distinct couplings that can arise. One, the \emph{lens-lens} coupling, accounts for how the change in the ray bundle shape by one lensing event affects the amount of lensing generated by a later lensing event: for example, convergence cause by one lens reduces the beam cross-section, and hence the differential deflection experienced during a subsequent lensing. There are also \emph{ray-deflection} effects, from changing gravitational potentials in the direction in which the ray path is bent. In this section we outline the calculation of the post-Born corrections to third order in $\Psi$. For post-Born corrections to the convergence power spectrum, contributions arise from couplings between the first and third order terms (``13''), and second-second terns squared (``22''). The couplings between the first and second order terms do not contribute to the power spectrum in the Limber approximation, even with non-zero potential bispectrum~\cite{Krause:2009yr}. However couplings of first and second order terms will be important in the discussion of the bispectrum in Sec.~\ref{sec:PBBispectrum}.

In order to introduce higher order corrections, we can introduce a Taylor series expansion of the potential $\Psi$ about an unperturbed geodesic position $\bfx_0$:
\begin{align}
\Psi (\bfx_0 + \delta \bfx ) \approx \Psi (\bfx_0) + \Psi_{,a} (\bfx_0) \delta x_a + \frac{1}{2} \Psi_{,ab} (\bfx_0) \delta x_a \delta x_b + \mathcal{O} (\Psi^4).
\end{align}
The last term is effectively third order in the potential but it must be retained as it will couple to the first order terms in the power spectrum. The deflection path is expanded perturbatively and implicitly depends on the higher order corrections to the potential
\begin{align}
\delta x_a &= \delta x_a^{(1)} + \delta x^{(2)}_a + \mathcal{O}(\Psi^3) , \\
\delta x_a^{(1)} &= -2 \int^{\chi}_0 d \chi^{\prime} \, W(\chi^{\prime} , \chi) \, \chi \chip \, \Psi_{,a} (\bft, \chip ) , \\
\delta x_a^{(2)} &= - 2 \int^{\chi}_0 d \chi^{\prime} \, W(\chi^{\prime} , \chi) \, \chi \chip  \, \Psi_{,ab} (\bft , \chi^{\prime} ) \, \delta x_b^{(1)} (\bft , \chi^{\prime} ) .
\end{align}
\n
Using these expressions, we can now express the potential $\Psi (\bfx)$ at the deflected position using integrals over the undeflected path
\begin{align}
\Psi (\bfx ) &= \Psi (\bfx_0 ) + \left( \delta x^{(1)}_a + \delta x^{(2)}_a \right) \Psi_{,a} (\bfx_0 ) + \frac{1}{2} \delta x^{(1)}_a \delta x^{(1)}_b \, \Psi_{,ab} (\bfx_0 ) + \mathcal{O} (\Psi^4) .
\end{align}
\n
This can be substituted into the expression for the deformation tensor $\psi_{ab}$ in order to derive a perturbative expansion for the deformation tensor to $\mathcal{O}(\Psi^4)$ \cite{Schneider:1997ge}
%\AL{citation or explanation}
\begin{align}
\psi_{ab} (\bft, \chi) &= 2 \int^{\chi}_0 d \chip \, \chi^{\prime 2} \, W(\chip,\chi) \, \Psi_{,ac} (\bfx^{\prime}) \, \calA_{b}^{c} (\bft,\chip), \\
&= 2 \int^{\chi}_0 d \chip \, \chi^{\prime 2} \, W(\chip,\chi) \, \Psi_{,ac} (\bfx^{\prime}) \, \left[ \delta_{b}^{c} - \psi_{b}^{c} (\bft,\chip) \right],
\label{eqn:LensingEqn}
\end{align}
\n
where $\bfx^{\prime}$ is the perturbed $\bfx$ at radial distance $\chip$. This equation can be solved iteratively order by order with the zeroth order solution being
\begin{align}
\left( \psi^a_b \right)^{(0)} &= 0.
\end{align}
\n
Working to $\mathcal{O}(n)$ is achieved iteratively allowing us to relate the deformation tensor to the angular deflection order by order \footnote{We adopt the notation of \cite{Fanizza:2015swa} for the iterative solutions.}
\begin{align}
\left( \psi^a_b \right)^{(n)} &= - \frac{\partial \theta^{a (n)}_{\calS}}{\partial \theta^b_{}} , \qquad n \geq 1 .
\end{align}
\n
Following \cite{Cooray:2002mj} we will also decompose the corrections to the deformation tensor into various contributions
\begin{align}
\psi^{(\alpha)}_{ab} &= +2 \int^{\chi_s}_0 d \chi \, \chi^2 \, W(\chi , \chi_s ) \, S^{(\alpha)}_{ab} ,
\end{align}
\n
where $\alpha$ denotes the term under consideration. The basic structure that we see in the higher order corrections can be described in terms of ray-deflection corrections (D), arising from corrections to the path of the photon, Lens-Lens coupling (L), arising from the iterative correction of the deformation tensor of distant lenses due to foreground lenses along the line of sight, and Deflection-Lens coupling (X) terms that arise as cross-terms between these two effects.
\n
\subsubsection{First Order}
At first order we find
\begin{align}
\left( \psi_{ab}  \right)^{(1)} &= 2 \int^{\chi_s}_0 d \chi \, \chi^2 \, W(\chi , \chi_s) \, \Psi_{,ab} (\chi) ,
\end{align}
\n
where we now drop the explicit $\bft$ dependence as terms are evaluated at the unperturbed position unless explicitly stated. The only source term at this order is given by $\Psi_{,ab}$.

\subsubsection{Second Order}
At second order the deformation tensor is given by
\begin{align}
\left( \psi_{ab}  \right)^{(2)} &= 2 \int^{\chi_s}_0 d \chi^{} \, \chi^2 \, W(\chi^{} , \chi_s) \, \left[ - \Psi_{,ac} (\chi) \psi^{(1)}_{cb} (\chi) + \Psi_{,acd} (\chi) \, \delta x_d^{(1)} (\chi^{}) \right] .
\label{eq:second_order}
\end{align}
\n
The two expanded source terms that arise at this order are given in Eqs.~\eqref{eqn:S2L} and \eqref{eqn:S2B} \cite{Cooray:2002mj,Shapiro:2006em,Krause:2009yr}, and describe respectively lens-lens coupling and a contribution from changing gravitational gradients in the direction of the ray deflection
(we call terms depending on $\delta x_d$, `ray-deflection' terms).

\subsubsection{Third Order}
Finally, at third order we find
\begin{align}
\label{eqn:def_tensor_3rd}
\left( \psi_{ab}  \right)^{(3)} &= 2 \int^{\chi_s}_0 d \chi^{} \, \chi^2 \, W(\chi^{} , \chi_s) \Bigg[ - \psi_{cb}^{(1)} \delta x^{(1)}_g (\chi) \Psi_{,acg} (\chi) - \psi_{cb}^{(2)}  \Psi_{,ac} (\chi)  \\
&\qquad \qquad \qquad \qquad \qquad \qquad \qquad \nonumber + \Psi_{,abg} (\chi) \, \delta x^{(2)}_g (\chi) + \frac{1}{2} \Psi_{,abgf} (\chi) \, \delta x^{(1)}_g (\chi^{}) \, \delta x^{(1)}_f (\chi^{})  \Bigg] .
\end{align}
\n
The source terms at third order are given by \ref{eqn:S3B}, \ref{eqn:S3X}, \ref{eqn:S3L}, \ref{eqn:S3X1} and \ref{eqn:S3X2} \cite{Cooray:2002mj,Shapiro:2006em,Krause:2009yr}. These results also agree with those derived in the geodesic light cone gauge (GLCG) when subleading terms, containing only two angular derivatives acting on the metric perturbations, are neglected \cite{Fanizza:2015swa}.

Note that the expressions here differ from those given in Eq.~(16) of Ref. \cite{Hagstotz:2014qea}.

Although the third-order contributions look complicated, in the Limber approximation the only non-vanishing contribution to the post-Born
power spectrum corrections is sourced by $\Psi_{,abgf} \, \delta x^{(1)}_g \, \delta x^{(1)}_f$, all other ``13" terms vanish. See Appendix~\ref{app:PBA} and Appendix~\ref{app:PBC} for further details.

\section{Lensing power spectra}
\label{sec:lenspower}
Correct expressions for the power spectra have been derived in Refs.~\cite{Hirata:2003ka,Krause:2009yr}, but since there has been
confusion in the literature, we review the arguments again in detail in Appendix \ref{appendix:power}. Since even with post-Born corrections the
lensing effect is still entirely described by a deflection angle, there are only two degrees of freedom: the E-mode (gradient) shear and convergence spectra remain equivalent, as do the field rotation and B-mode (curl) shear.

 In the context of CMB lensing, Ref.~\cite{Su:2014mga} directly solve the Boltzmann equation using a diagrammatic approach to include interaction terms and non-trivial correlations to arbitrarily high orders. They find that the corrections to the CMB temperature power spectrum from lens-lens couplings are on the order of $\lesssim 0.1\%$ of the CMB temperature power spectrum for $\ell$ up to $3000$. More recently, \cite{Hagstotz:2014qea} claimed that post-Born corrections lead to large effects on the CMB lensing spectra with contributions that are comparable to the power generated by nonlinear structure formation on small scales. They claim that post-Born corrections lead to relative changes on the order of $\sim 1\%$ for the E-mode spectrum and $\sim 10\%$ for the B-mode spectrum. In the next few sections we detail the derivation of the post-Born corrections and provide heuristic arguments as to why the corrections are small, in agreement with numerical simulations incorporating post-Born effects \cite{Calabrese:2014gla}. Finally, we note that we have neglected contributions from tensor modes, large scale structure vorticity, non-lensing GR effects and second order vectors sourced by scalars~ \cite{Mollerach:1997up,Hu:2001yq,Mollerach:2003nq,Cooray:2005hm,Li:2006si,Bernardeau:2009bm,Dai:2013nda,Padmanabhan:2013xfa,Namikawa:2014lla,Saga:2015apa,Adamek:2015mna}.

\subsection{Post-Born Corrections to $C^{\kappa \kappa}_{\ell}$}

\begin{figure*}
\begin{center}
\includegraphics[width=140mm]{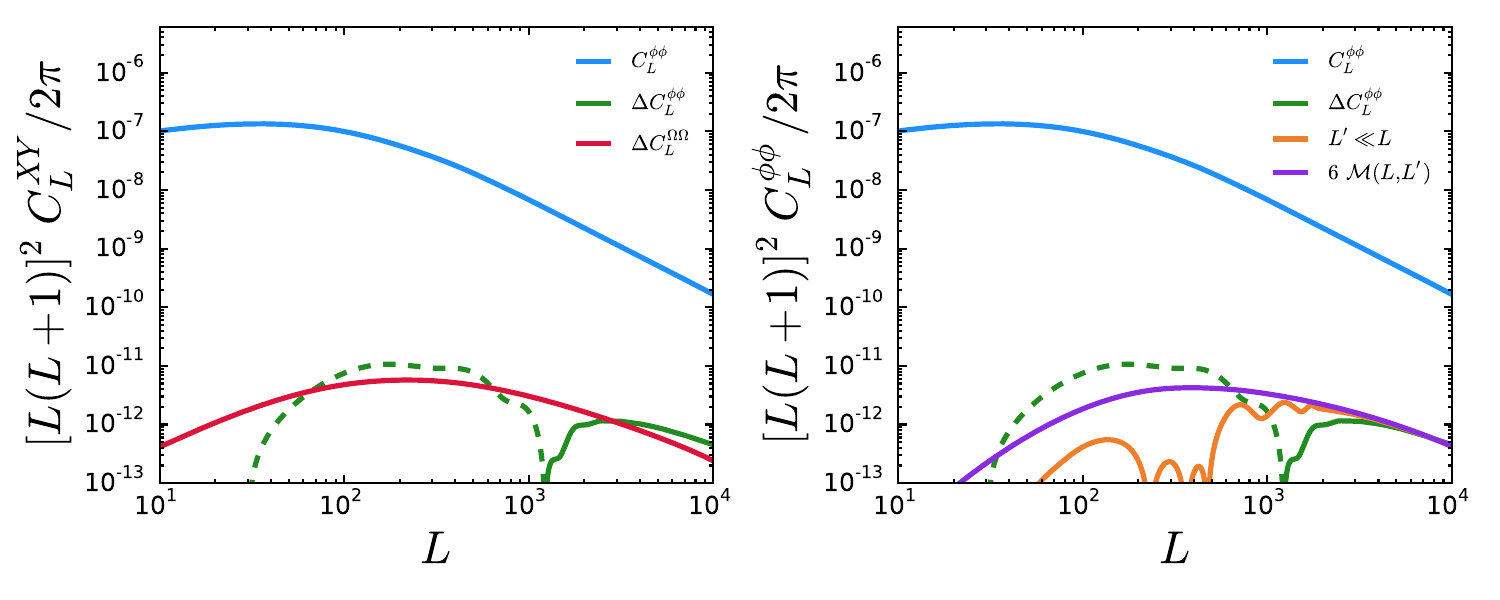}
\caption{
The linear-order lensing potential (convergence) power spectrum (blue), and the post-Born corrections to it (green).
At low $L$ the post-Born corrections to the convergence are negative (dashed); at $L \agt 1000$ the corrections become positive. The lensing potential spectrum with and without post-Born corrections are indistinguishable by eye.
The left plot also shows the post-Born curl (rotation) spectrum (red).
The right plot shows how the high-$L$ form of the convergence post-Born corrections are approximately fit by the coupling to larger scales, showing the $L^{\prime} \ll L$ series expansion approximation in Eq.~\eqref{eq:curlm_series} (orange) and the smooth, dominant $6 \mathcal{M}(L,L^{\prime})$ contribution from this series expansion (purple).
}
\label{fig:Clkk_All_PB}
\end{center}
\end{figure*}

The leading correction to the convergence spectrum reduces to
\begin{align}
\Delta C^{\kappa\kappa}_{L} &= 4 \int \frac{d^2 \vL'}{(2 \pi)^2} \, \left( \frac{(\vL \cdot \vL'')^2 (\vL'' \cdot \vL')^2}{L^{\prime \prime 4} L^{\prime 4}} \Mkappa (L'',L^{\prime}) - \frac{(\vL \cdot \vL')^2}{L^{\prime 4}} \Mkappa(L^{\prime},L) \right),
\label{eq:totcurlyM}
\end{align}
where $\vL'' =\vL-\vL'$ and the coupling matrix is given by
\begin{align}
{\Mkappa}(L ,L^{\prime}) &\equiv L^4 \, L^{\prime 4} \, M(L , L^{\prime}) , \nonumber\\
&= L^4 \int^{\chi_s}_0 d \chi \, \frac{W^2 (\chi,\chi_s)}{\chi^2} \, P_{\Psi \Psi} \left( \frac{L}{\chi} , z(\chi) \right) \, C^{\kappa \kappa}_{L^{\prime}}(\chi).
\label{eq:curlyM}
\end{align}
The coupling matrix is of the form of the linear result for the convergence power spectrum with additional weighting by $C^{\kappa \kappa}_{L^{\prime}}(\chi)$, the convergence power to radial distance $\chi$. Parametrically
\be
\Mkappa(L ,L^{\prime}) \sim \clo\left( C^{\kappa \kappa}_{L}(\chi_s) C^{\kappa \kappa}_{L^{\prime}}(\bar\chi )\right)
\ee
where $\bar{\chi}$ is some effective average along the line of sight ($\bar{\chi}\sim \chi_s/2)$.

\begin{figure*}
\begin{center}
\includegraphics[width=140mm]{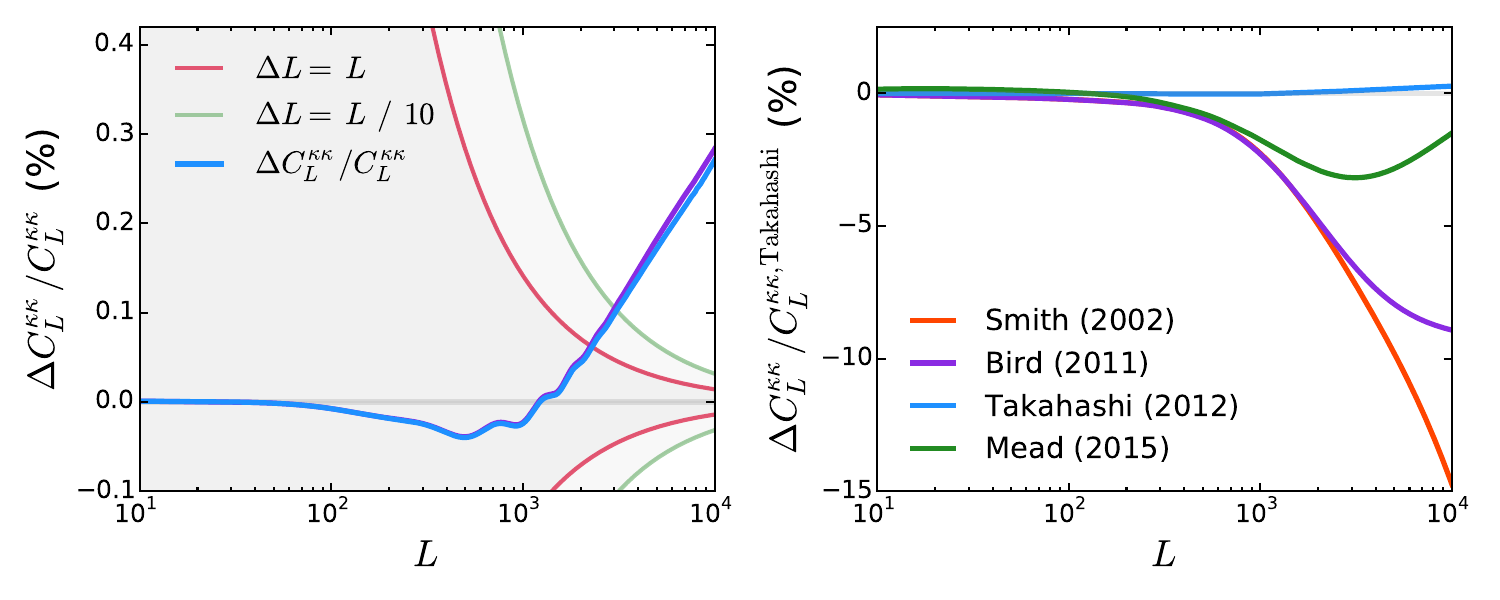}
\caption{
The left plot shows the fractional contribution of the post-Born corrections to the convergence power spectrum $C^{\kappa\kappa}_{L}$ as a percentage. Cosmic variance bands show $\pm \sqrt{2C^2/((2L+1)\Delta L)}$ for $\Delta L = L$ and $\Delta L = L/10$. At $L \sim 10^4$, the contributions are on the order $\sim 0.25$\%. For scales accessible by a number of upcoming CMB missions, i.e. $L \sim 5000$, the contributions are on the order of $0.2$\% with the post-Born corrections dominating over cosmic variance at $L \sim 3000$ for $\Delta L = L / 10$. The right plot demonstrates the changes in the convergence power spectrum due to different nonlinear power spectrum fits: it shows changes in the total power spectrum (including post-Born corrections) generated using variations of \HALOFIT\ from
Refs.~\cite{Smith:2002dz} (purple, original),
\cite{Bird:2012MNRAS.420.2551B} (purple), \cite{Mead:2015yca} (green, most recent), compared the reference result without post-Born corrections from Ref.~\cite{Takahashi:2012em} (blue). The post-Born corrections are a subdominant error in comparison to current non-linear modelling uncertainty. }
\label{fig:Clkk_PB_Error}
\end{center}
\end{figure*}

The two terms in Eq.~\eqref{eq:totcurlyM} correspond to the ``22'' and ``13'' contributions to the power spectrum from
the perturbative expansion. In a statically isotropic universe the lensing correlation function is invariant under constant angular displacements, but this symmetry is not reflected in the perturbative expansion of the fields (which can change by $\clo(1)$ when the wavelength is smaller than the deflection). The symmetry manifests itself at the power spectrum level as a cancellation between large ``22'' and ``13'' contributions to the power spectra from large scales, in the same way as in many other standard perturbation theory analyses in cosmology\footnote{Numerical problems with evaluating these contributions probably explain the discrepant results of Ref.~\cite{Hagstotz:2014qea}. We control the numerical calculation by differencing the two terms inside the integration, using the series expansion of Eq.~\eqref{eq:curlm_series} for $L'\ll L$, or re-writing the integral in the way suggested by~\cite{Krause:2009yr}, with excellent agreement between the different methods.}. Nonetheless, the small-scale post-Born correction is dominated by the residual coupling to larger-scale modes.
Contributions from large-scales appear in the integrand of Eq.~\eqref{eq:curlyM} where $L'\ll L$ or $L''\ll L$,
and performing a series expansion gives the leading order result
\begin{align}
\Delta C^{\kappa \kappa}_{L} & \sim \frac{1}{8 \pi} \int_{L'\ll L} L^{\prime} d L^{\prime} \left\{\left( 2 \frac{\partial \ln \Mkappa (L ,L')}{\partial \ln L} + 3 \left[ \frac{\partial^2 \ln \Mkappa (L ,L^{\prime})}{\partial \ln L^2} + \left( \frac{\partial \ln \Mkappa (L ,L')}{\partial \ln L} \right)^2 \right] - 2 \right) \Mkappa (L , L^{\prime})
+ 6\Mkappa (L^{\prime} , L) \right\}.
\label{eq:curlm_series}
\end{align}
The two terms are the same order, but the second term numerically dominates and qualitatively describes the shape and sign of the post-Born power spectrum at $L \gg 1000$.
Parametrically we then have
\begin{align}
\Delta C^{\kappa \kappa}_{L} &\sim
\frac{6}{8 \pi}\int^L d\log L' L'^2 \Mkappa(L,L')  \sim
C^{\kappa \kappa}_{L} \frac{6}{8 \pi}\int^L d\log L'\, L'^2  C^{\kappa \kappa}_{L^{\prime}}(\bar\chi ) \\&\sim \clo\left(C^{\kappa \kappa}_{L} \left\langle [\bar\kappa(\bar\chi)]^2\right\rangle\right).
\end{align}
Thus the small-scale lensing power is increased by $\left\langle [\bar\kappa(\bar\chi)]^2\right\rangle$, the variance of the convergence due to lensing modes on larger scales. Since the lensing variance from modes $L \alt 2000$ most relevant for CMB lensing is less than $0.005$, this is a very small correction. The fractional contribution is in reality further suppressed by negative contributions from similar and smaller scales, and because non-linear growth of structure boosts the leading-order convergence power spectrum and reduces the effective path length for post-Born contributions because relatively more of the lensing happens at low redshift.

On scales $L \ll 1000$ the post-Born spectrum is very small but dominated by negative contributions. The negative contributions mostly come from coupling of similar scales, but there is a tail coupling to smaller scales which can be approximated by the series expansion result for $L'\gg L$:
\begin{multline}
\Delta C_L^{\kappa\kappa} \sim \frac{L^2}{\pi}\int_{L'\gg L} \frac{d L'}{L'} \biggl\{ \Mkappa(L',L')-\Mkappa(L,L')
\\
+\frac{L^2}{8L'^2} \left(  2 \frac{d\ln \Mkappa(L',L')}{d \log L'} + \left[\frac{d^2\ln \Mkappa(L',L')}{d \,{\log L'}^2} +
\left(\frac{d\ln \Mkappa(L',L')}{d \log L'}\right)^2\right] - 2\right)
\biggr\},
\end{multline}
where derivatives are with respect to the first argument. This expression captures the negative sign and shape of the post-Born contributions on large scales, but not the amplitude since couplings to smaller scales do not dominate.
There is a change in sign of the total contribution around $L\sim 1000$, since on smaller scales the positive coupling to larger scales (Eq.~\eqref{eq:curlm_series}) starts to dominate.

Numerical results for the post-Born correction are shown in Fig.~\ref{fig:Clkk_All_PB}, compared to the asymptotic result of Eq.~\eqref{eq:curlm_series} and its dominant sub-term. The series expansion result is a good fit to the asymptotic small-scale correction. On all scales the post-Born contribution is very small. Fig~\ref{fig:Clkk_PB_Error} shows that the $\alt 0.2\%$ contribution to the convergence power is well below cosmic variance at $L\alt 3000$, and highly subdominant to modelling uncertainties that arise from non-linear structure growth.

\subsection{Field rotation spectrum $C^{\omega \omega}_L$}

Since there is no first order field rotation, its power spectrum comes entirely from second-order lens-lens coupling terms, and is given by~\cite{Cooray:2002mj,Hirata:2003ka}
\be
C_L^{\omega\omega} = 4\int \frac{d^2 \vL'}{(2\pi)^2} \frac{ (\vL\times \vL')^2 (\vL'\cdot \vL'')^2}{{L'}^4(L'')^4} \Mkappa(L',L'').
\ee
The numerical calculation is shown in Fig.~\ref{fig:Clkk_All_PB}, and is of comparable size to the post-Born corrections to the gradient spectrum. However, since the rotation spectrum is zero at leading order, there is no leading cosmic variance contribution to limit detectability, making the rotation signal potentially more detectable.
The curl lensing signal can be measured using maximum-likelihood or quadratic estimators, as for the gradient signal~\cite{Cooray:2005hm}.
However the simplest quadratic estimator for $\omega$ is not actually exactly orthogonal to the convergence estimator, and there are also non-zero $N^{(1)}_L$ contributions to the signal from the much larger convergence field. Here for a first estimate of observational relevance we simply assume that these effects can be removed accurately (for example the $N^{(1)}_L$ can be estimated from the realization of the large-scale well-resolved convergence modes, or the convergence field can be delensed).
Table~\ref{table:sigmas} shows however that even under ideal assumptions the curl power spectrum signal will remain undetectable until noise levels reach $\sim 0.25\muKarcmin$. Note that the rotation modes are uncorrelated to the leading-order convergence modes, meaning that the cross-spectrum vanishes $C^{\kappa \omega}_L = 0$, as expected from their different parity.

\section{Post-Born and Total Bispectrum}
\label{sec:PBBispectrum}

Although the post-Born effects are hard to see in the auto power spectra, a detailed measurement of large-scale structure could in principle be used to predict what the post-Born corrections should be, and hence serve as a template for seeing the post-Born effects in cross-correlation with a CMB lensing reconstruction. Since the leading post-Born effect is second order, the template would be a quadratic combination of the large-scale density fields. The correlation can also be described in terms of a bispectrum between a pair of tracers of the large-scale structure and a CMB lensing field.

For any three fields the bispectrum is defined by
\begin{align}
\langle X (\vL_1) Y (\vL_2) Z (\vL_3) \rangle &= (2 \pi)^2 \delta_D (\vL_1 + \vL_2 + \vL_3 ) \, b_{L_1,L_2,L_3}^{XYZ}.
\end{align}
We focus here on the case where all three tracers are from CMB lensing. The leading contribution to the bispectrum is given by correlation of second-order post-Born corrections with two linear terms.  Expanding Eq.~\eqref{eq:second_order} in harmonics gives the second-order term
\be
\psi_{ab}^{(2)}(\vL) = -4 \int_0^{\chi_s}  d\chi W(\chi,\chi_s)\int_0^\chi d\chi' W(\chi',\chi) \int \frac{d^2 {\vL'}}{(2\pi)^2} \,\vL_a'\vL_b \, \vL'\cdot(\vL-\vL') \Psi(\vL',\chi)\Psi(\vL-\vL',\chi'),
\label{eq:distortiontwo}
\ee
from which we can form the convergence contributions from $\kappa = \frac{1}{2}\delta_{ab}\psi_{ab}$ and rotation contributions from
$\omega = \frac{1}{2}\epsilon_{ab}\psi_{ab}$.
Contracting with the linear convergence then gives
\begin{align}
b^{\kappa^{(1)}\kappa^{(1)} Z^{(2)} }_{L_1 L_2 L_3}  &=  2 L^2_1 L^2_2 \, \left[ \vL_1 \cdot \vL_2 \right] \int_0^{\chi_s} d\chi \frac{W(\chi,\chi_s)^2}{\chi^2} \int_0^{\chi}d\chi' \frac{W(\chi',\chi)W(\chi',\chi_s)}{{\chi'}^2} \nonumber \\
&\qquad \qquad \qquad \qquad \qquad \times \left[  \left( \vL_1 \otimes \vL_3 \right) P_{\Psi\Psi}\left(\frac{L_1}{\chi},z(\chi)\right) P_{\Psi\Psi}\left(\frac{L_2}{\chi'},z(\chi')\right) + (L_1 \leftrightarrow L_2) \right]\\
&=
 2 \frac{\vL_1 \cdot \vL_2 }{L_1^2 L_2^2} \left[\vL_1 \otimes \vL_3 \Mkappa_s(L_1,L_2) + \vL_2 \otimes \vL_3 \Mkappa_s(L_2,L_1)\right],
\label{eq:kkz}
\end{align}
where $\otimes$ is a dot product for $Z=\kappa$ and a cross product for $Z=\omega$, and we defined
\be
\Mkappa_s(L,L') \equiv L^4\int_0^{\chi_s}d\chi \frac{W(\chi,\chi_s)^2}{\chi^2} P_{\Psi\Psi}\left(\frac{L}{\chi}, z(\chi)\right) C_{L'}^\kappa(\chi,\chi_s).
\ee

\subsection{The $\kappa\kappa\kappa$ bispectrum}

\begin{figure*}
\begin{center}
\includegraphics[width=180mm]{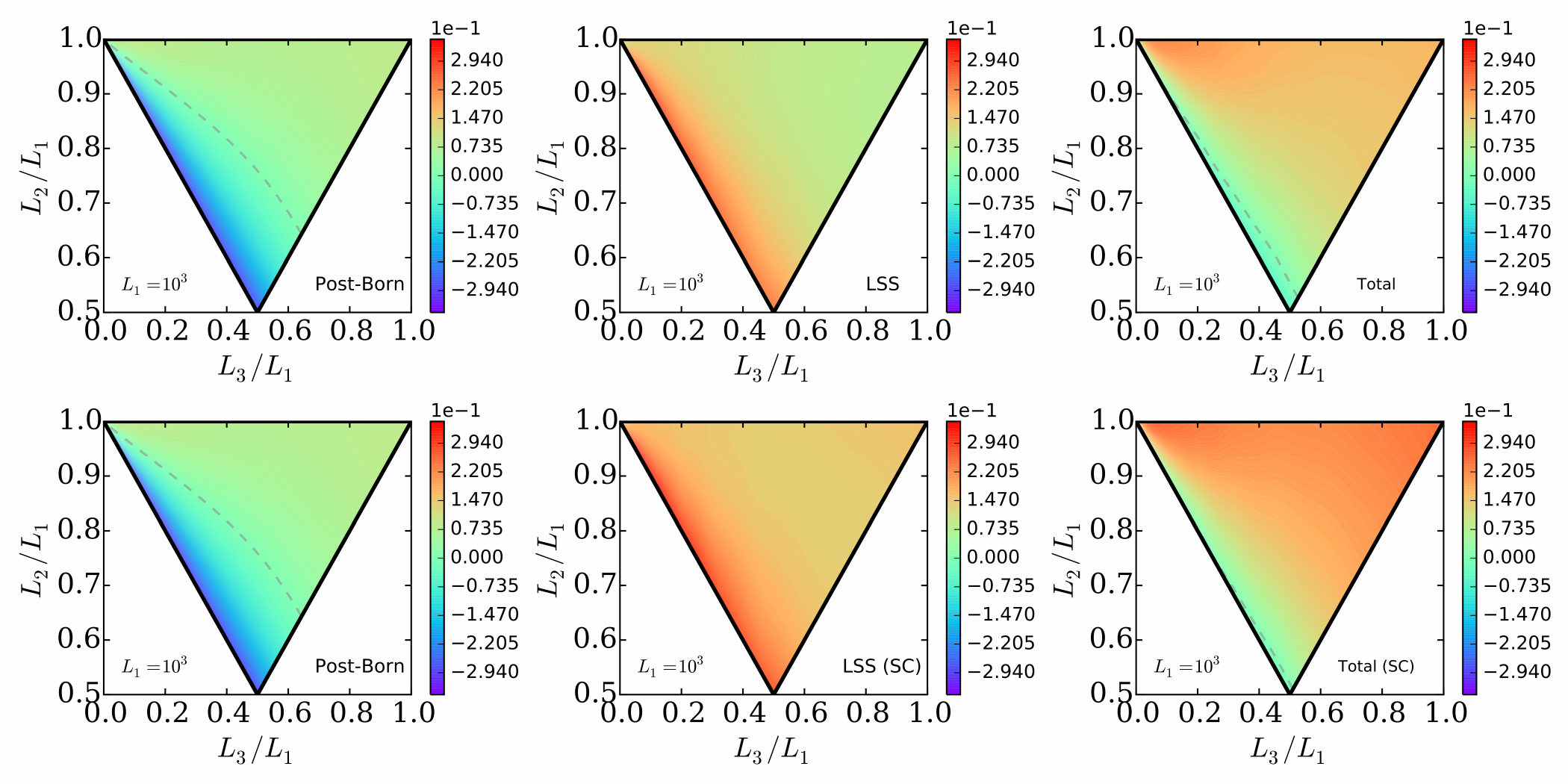}
\caption{
{
Slices through the weighted convergence bispectrum $(L_2 L_3)^{1/2} \, b^{\kappa\kappa\kappa}_{L_1 L_2 L_3} / (C^{\kappa\kappa}_{L_1} C^{\kappa\kappa}_{L_2} C^{\kappa\kappa}_{L_3} )^{1/2}$ for $L_1 = 10^3$. The top row shows the tree-level LSS bispectrum and the bottom row shows the non-linear fit of \citet{Scoccimarro:2000ee} (``SC''). The left plots show the post-Born contributions, the middle plots the large-scale structure contributions and the right plots the cancellations that occur due to negative contributions from the post-Born bispectrum in approximately flattened configurations, i.e. $L_1 + L_2 \sim L_3$. For approximately equilateral configurations, i.e. $L_1 \sim L_2 \sim L_3$, we find an enhancement of the total bispectrum. The grey dashed line denotes the $b^{\kappa\kappa\kappa}=0$ contour. In the equilateral limit, the tree-level LSS bispectrum is enhanced by a factor of $\sim 2$ by the post-Born corrections and the non-linear SC LSS bispectrum by a factor of $\sim 1.5$.
}
}
\label{fig:bispectrum_kkk_PB_LSS_triangle}
\end{center}
\end{figure*}

For the convergence the total post-Born bispectrum follows immediately from Eq.~\eqref{eq:kkz} as
\be
\label{eqn:kkk_PB}
{b}^{\kappa\kappa\kappa}_{L_1 L_2 L_3} = 2\frac{\vL_1\cdot\vL_2}{L_1^2L_2^2}\left[ \vL_1\cdot \vL_3 \Mkappa_s(L_1,L_2)
 +\vL_2\cdot \vL_3 \Mkappa_s(L_2,L_1)\right] + \text{cyc. perm.}
 \qquad {[\text{Post-Born}]}
\ee
This agrees with the galaxy lensing result of Ref.~\cite{Dodelson:2005rf} when restricted to observing the convergence (trace of the distortion tensor) directly.

Non-linear structure growth will also give additional contributions from the large-scale structure bispectrum,
which in the Limber approximation has the form~\cite{Takada:2003ef}
\be
b^{\kappa\kappa\kappa}_{L_1 L_2 L_3} = L_1^2 L_2^2 L_3^2\int_0^{\chi_s} d\chi \frac{W(\chi,\chi_s)^3}{\chi^4}B^{\Psi\Psi\Psi}(L_1/\chi, L_2/\chi, L_3/\chi;\chi).
\qquad {[\text{LSS}]}
\ee
The potential bispectrum on the right-hand side of this equation can be approximated by the tree-level result for the fractional matter density perturbations~\cite{Bernardeau:2001qr}:
\be
b^{\delta\delta\delta}(k_1, k_2, k_3;\chi) = 2F_2(\vk_1,\vk_2;z) P_{\delta\delta}(k_1,z(\chi))P_{\delta\delta}(k_2,z(\chi)) + \text{cyc. perm.} ,
\label{eq:bdelta}
\ee
where
\be
F_2(\vk_1,\vk_2;z) = \frac{5}{7}A(k_1,k_2;z) + B(k_1,k_2;z)\frac{\vk_1\cdot\vk_2}{2k_1k_2}\left(\frac{k_1}{k_2}+\frac{k_2}{k_1}\right)
+ C(k_1,k_2;z)\frac{2}{7} \frac{(\vk_1\cdot\vk_2)^2}{k_1^2 k_2^2}
\label{eq:Ftwo}
\ee
and $k^2\Psi(\vk,z)  \approx -\gamma(z)\delta(\vk,z)$ with $\gamma(z)\approx k^2\sqrt{P_{\Psi\Psi}(k,z)/P_{\delta\delta}(k,z)}$ approximately independent of $k$.
The baseline tree-level result has $A=B=C=1$ in Eq.~\eqref{eq:Ftwo}, so that $F_2$ is independent of redshift, but even in this case we use the non-linear $P_{\delta\delta}$ from \HALOFIT\ to improve accuracy~\cite{Scoccimarro:2000ee}. We also consider
extended fitting formulae for $A$, $B$ and $C$ from ~\citet{Scoccimarro:2000ee} (denoted ``SC'') to assess the order of magnitude of fully non-linear corrections beyond tree level\footnote{We use a dewiggled form for $n\equiv d\ln P_{\delta\delta,\rm{lin}}/d\ln k$ in the fitting function following Ref.~\cite{GilMarin:2011ik}. Ref.~\cite{GilMarin:2011ik} (``GM'') also provide updated fits, but they are not validated at high redshift and may behave in an unphysical way there, so we restrict to the original fit of Ref.~\cite{Scoccimarro:2000ee}. Using the updated fits would slightly increase the LSS bispectrum signal, but not change results qualitatively. At $L=1000$ the equilateral lensing bispectrum is enhanced by $\approx 2\times$ compared to tree level using the SC fit, and $\approx 2.5\times$ in the GM fit, but the difference is much less for folded configurations (where the non-linear enhancement at $L_1=1000$ is less than 20\%).
}

\begin{figure*}
\begin{center}
\includegraphics[width=180mm]{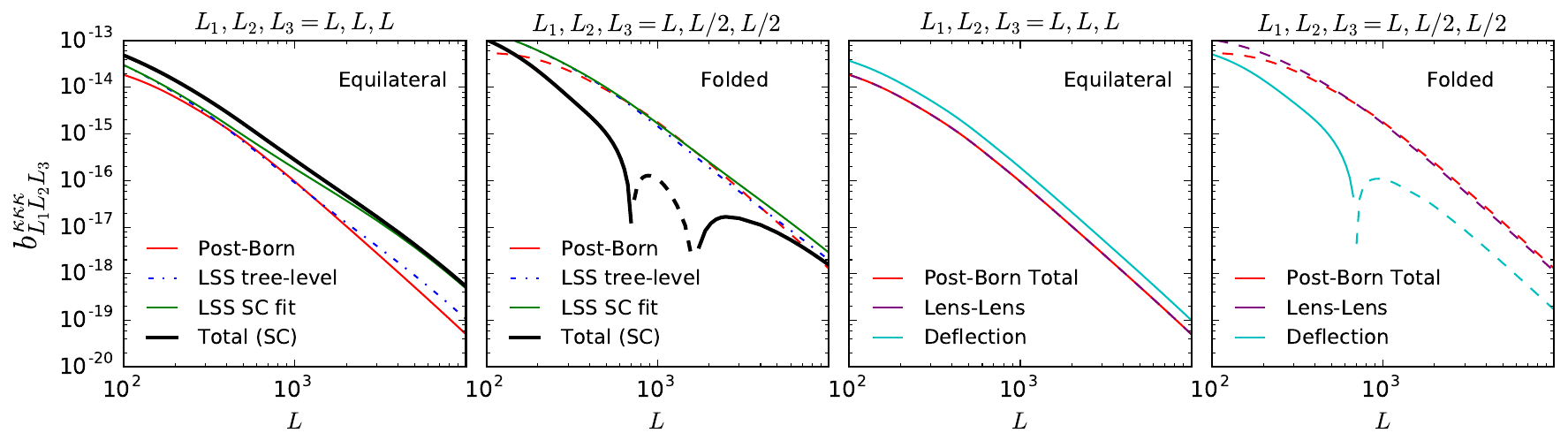}
\caption{
Left two plots: comparison between slices of the lensing convergence bispectrum from post-Born corrections (red) and LSS (tree-level with and without the non-linear fit of SC~\cite{Scoccimarro:2000ee}) for equilateral and folded (mode-aligned) configurations. The total (black, post-Born plus LSS with SC) is suppressed by cancellations for the folded shape, and slightly enhanced for equilateral. The right two plots show the contributions to the post-Born bispectrum from the lens-lens term (purple) and ray deflection (cyan): for aligned folded shapes the lens-lens effect is substantially larger and negative; for the equilateral configuration the ray-deflection term is twice as large as lens-lens term, giving a positive total equal to minus the lens-lens contribution. Negative contributions are shown dashed.
}
\label{fig:bispectrum_kkk_PB_LSS}
\end{center}
\end{figure*}

The different shapes of the LSS and post-Born bispectra are illustrated in the slices shown in Figs.~\ref{fig:bispectrum_kkk_PB_LSS_triangle} and~\ref{fig:bispectrum_kkk_PB_LSS}. The post-Born signal cancels a significant part of the LSS bispectrum, and slightly enhances the equilateral signal, substantially changing the total shape from that expected from LSS alone.

\begin{figure*}
\begin{center}
\includegraphics[width=180mm]{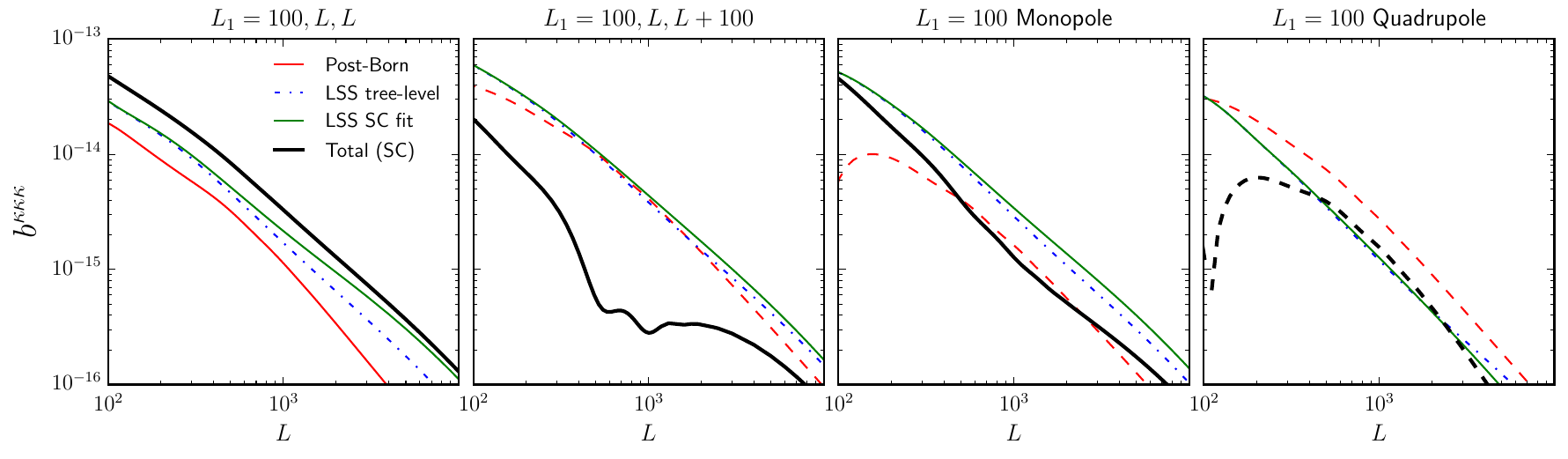}
\caption{
Lensing convergence bispectrum with largest mode $L_1=100$. The leftmost pair of plots show slices
for perpendicular and parallel large and small-scale modes; the rightmost pair of plots show the decomposition
into monopole and quadrupolar angular dependence on the mode orientation. Red lines show the post-Born contributions, where dashed lines are negative; the large-scale structure bispectrum (LSS) is shown for the tree-level result (blue dot-dash) and the approximate numerical fit of Ref.~\cite{Scoccimarro:2000ee} (green). The total of the post-Born and LSS fit (thick black) is suppressed by cancellations for aligned modes and the monopole. The total quadrupole is dominated by the post-Born contribution on small scales and negative (dashed).
}
\label{fig:squeezed_kappa}
\end{center}
\end{figure*}

Since lensing reconstruction noise grows rapidly on small scales, most of the signal in the bispectrum is in semi-squeezed shapes, involving one larger-scale mode.
 For $k_1\ll k$, with the short mode $\vk\equiv (\vk_2-\vk_3)/2$, we have
\be
b^{\delta\delta\delta}(k_1,k,\varphi_k) = \left( \frac{17}{7} - \frac{1}{2}\frac{d\ln P_{\delta\delta}(k)}{d\ln k}
+\cos(2 \varphi_k)\left[ \frac{4}{7}  - \frac{1}{2}\frac{d\ln P_{\delta\delta}(k)}{d\ln k} \right]
   \right)P_{\delta\delta}(k_1)P_{\delta\delta}(k) + \clo\left(\frac{k^2_1}{k^2}\right),
   \label{eq:lss_squeezed}
\ee
where $\varphi$ is the angle between $\vk_1$ and $\vk$, and on the scales of interest the derivatives are negative. If we consider the baseline tree-level bispectrum where $F_2$ is independent of redshift, we can also write
\be
b^{\kappa\kappa\kappa}_{L_1 L_2 L_3} = 2F_2(L_1,L_2) \Fkappa(L_1,L_2) + \text{cyc. perm.}
 \qquad {[\text{LSS}]}
\ee
where
\be
 \Fkappa(L_1,L_2)\equiv -\int_0^{\chi_s} d\chi \chi^2 \left[W(\chi,\chi_s)\gamma(z(\chi))\right]^3 P_{\delta\delta}(L_1/\chi,z(\chi))
  P_{\delta\delta}(L_2/\chi,z(\chi))  .
\ee
The squeezed structure of Eq.~\eqref{eq:lss_squeezed} is then inherited by the convergence bispectrum,
\be
b_{\text{LSS}}^{\kappa\kappa\kappa}(L_1,L,\varphi_L) = \left(\frac{17}{7} - \frac{1}{2}\frac{d\ln \Fkappa(L_1,L)}{d\ln L}
+\cos(2\varphi_L)\left[ \frac{4}{7}  - \frac{1}{2}\frac{d\ln \Fkappa(L_1,L)}{d\ln L} \right]
   \right) \Fkappa(L_1,L) + \clo\left(\frac{L^2_1}{L^2}\right),
   \label{eq:Fsqueezed}
\ee
where  $L_1\ll L$, with the short mode $\vL\equiv (\vL_2-\vL_3)/2$. In the post-Born case we have analogously
\be
b_{\text{Post-Born}}^{\kappa\kappa\kappa}(L_1,L,\varphi_L) = -2\Mkappa_s(L_1,L) + \frac{d \Mkappa_s(L,L_1)}{d\ln L} +
\left( -2\Mkappa_s(L,L_1)-2\Mkappa_s(L_1,L) + \frac{d \Mkappa_s(L,L_1)}{d\ln L}  \right)\cos 2\varphi_L   + \clo\left(\frac{L_1^2}{L^2}\right).
\ee
On the scales of interest the coefficients of the monopole and quadrupole part of the post-Born spectrum are both negative, but in Eq.~\eqref{eq:Fsqueezed} they are both positive. We can therefore expect a partial cancellation between the contributions to the bispectrum. Although the approximate results of Eqs.~\eqref{eq:lss_squeezed} and ~\eqref{eq:Fsqueezed} are based on a series expansion, since the next term is $\clo(L_1^2/L^2)$ they provide quite a good qualitative fit to the full result for $L\agt 2L_1$.

Whether there is total reduction depends on the particular triangle shape, since the relative size of the quadrupolar contribution is much larger in the post-Born case. In the particular case where the long and short modes are orthogonal ($\cos2\varphi=-1$) we have the post-Born contribution $b^{\kappa\kappa\kappa}\approx 2\Mkappa_s(L,L_1)$ which is positive, and LSS contribution
 $b^{\kappa\kappa\kappa} \approx\frac{13}{7}\Fkappa(L_1,L)$, also positive. However, when the modes are parallel ($\cos2\varphi =1$) they have opposite sign:
 $b^{\kappa\kappa\kappa}\approx  -4\Mkappa_s(L_1,L) -2\Mkappa_s(L,L_1) + 2\frac{d \Mkappa_s(L,L_1)}{d\ln L} $,
 $b^{\kappa\kappa\kappa}  \approx (3 - \frac{d\ln \Fkappa(L,L_1)}{d \ln L})\Fkappa(L,L_1)$.
 Contributions for $\cos2\varphi\sim 1$ are larger than $\cos 2\varphi\sim -1$ because the monopole and quadrupole parts contribute with the same sign, and hence the net effect is that the post-Born bispectrum reduces the total signal compared to that from the tree-level large-scale structure bispectrum alone.
This is illustrated by typical numerical results for semi-squeezed shapes are shown in Fig.~\ref{fig:squeezed_kappa} as a function of the small-scale mode.

\begin{figure*}
\begin{center}
\includegraphics[width=120mm]{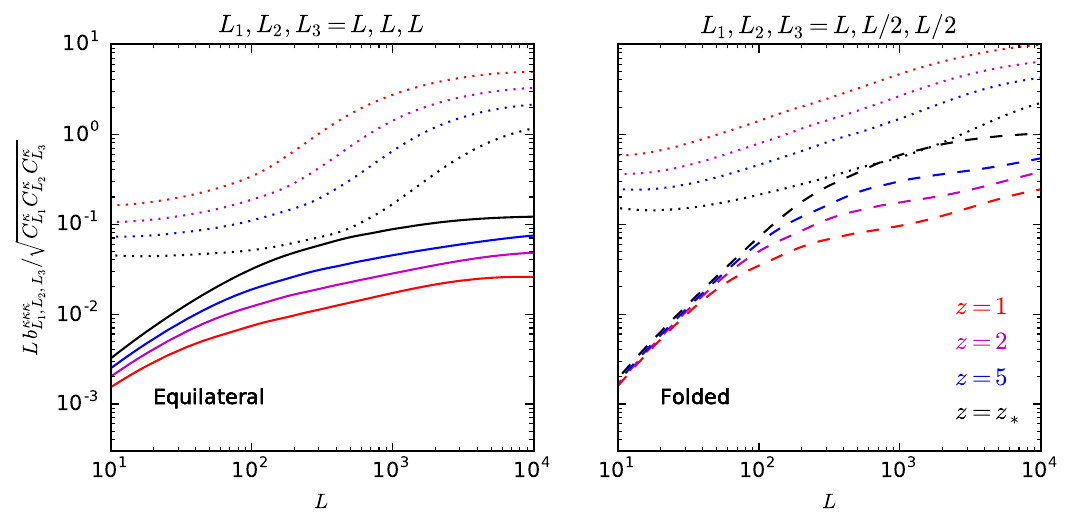}
\caption{
Scaling with redshift of the LSS (SC fit, dotted) and post-Born (solid, positive, and dashed, negative) contributions to the convergence bispectrum for equilateral and folded shapes. At low redshifts the post-Born contributions are much smaller than LSS, but for CMB lensing ($z=z_*$) they are coincidentally of comparable order of magnitude on relevant scales.
The bispectra here are plotted normalized by the convergence power to remove the total growth in lensing signal with redshift, so the curves represent a measure of the amount of non-Gaussianity.
}
\label{fig:redshift_scaling}
\end{center}
\end{figure*}

The shape of the bispectrum and partial cancellation with the LSS signal can easily be understood in simple cases.
A converging lens causes contraction of the ray bundle, so if there are two lens events the ray area is smaller at the second lensing event, the difference in potential gradients across the beam is lower, and hence there is less effect from the second lens than if the first lens had not been there: mathematically, $(1-\kappa)^2 > 1-2\kappa$, so combining two converging lenses has less effect than adding the convergences linearly.
An overdensity will have positive convergence, so consider combining a large lens and a small lens both of positive convergence: since the combined effect is smaller than obtained by linearly adding the two convergences, there is an anticorrelation between the large-scale lens convergence and the amplitude of the total small-scale convergence observed (corresponding to a negative bispectrum). On the other hand density perturbations grow faster the more dense they are, so there is a positive correlation between large overdensities and the magnitude of small scale perturbations on top of them (corresponding to a positive bispectrum).

The above argument applies in the case where all the modes are aligned, so that the contraction of the ray bundle in the two lensing event is in the same direction. More generally, there are two effects contributing to the post-Born bispectrum: From Eq.~\eqref{eq:second_order}, there is the lens-lens coupling (as described above), but there is also the additional distortion that arises if the distortion field is changing in direction in which the ray is deflected. For example, a ray passing the edge of an overdensity may be deflected towards the centre, where the potential gradients are larger, leading to more lensing that if the two contributions had been added independently.
In the case of the squeezed bispectrum, when the small and large-scale modes are orthogonal to each other the lens-lens coupling is effectively zero because the deflections are in opposite directions: the bispectrum is then dominated by the deflection effect which is positive, so there is a small enhancement of the LSS bispectrum. For mode-aligned squeezed shapes, the lens-lens effect is much larger, which means the dominant effect on the total bispectrum is a suppression of the LSS signal by post-Born contributions. The modes are also aligned in the folded shape: the post-Born effect also partly cancels with the large-scale structure signal
 that arises from the non-linear pancaking as structure grows and the formation of filaments\footnote{In 3D folded corresponds to planar overdensities, equilateral to filaments~\cite{Lewis:2011au,Schmittfull:2012hq}. For lensing we only see things in projection, so folded corresponds to planes aligned along the line of sight and filaments perpendicular to the line of sight; equilateral to filaments parallel to the line of sight and clusters.}. At lower redshifts, the non-linear growth of filaments and clusters leads to a large equilateral LSS bispectrum; here the post-Born effect is actually positive, since the ray-deflection effect is twice as large as the lens-lens suppression. However, for CMB lensing the redshifts being probed are relatively high, and very small-scales cannot be reconstructed, so equilateral shapes do dominate the signal from LSS alone.
The relative contributions of lens-lens and ray-deflection terms are shown in the right two plots of Fig.~\ref{fig:bispectrum_kkk_PB_LSS}.

 The fact that the post-Born and LSS contributions are of comparable order of magnitude for CMB lensing is a coincidence that only occurs for very high redshift lensing source planes, making post-Born contributions important to determine the total shape and amplitude rather than a small correction as for galaxy lensing at lower redshift~\cite{Dodelson:2005rf}. Fig.~\ref{fig:redshift_scaling} shows the scaling with redshift of the non-Gaussianity from post-Born and non-linear LSS for a couple of shapes. For higher-redshift source planes there are more independent lenses along the line-of-sight, a larger fraction of the path length is at higher redshifts where non-linear effects are smaller, and the low-redshift non-linear sources have lower weight in the lensing kernel: the non-Gaussianity from LSS gets substantially suppressed. However the lensing amplitude grows with redshift (see Fig.~\ref{fig:LEF}), so the impact of multiple-lens couplings increases, and the post-Born contributions have larger amplitude: post-Born contributions become relatively much more important at high redshift.

\begin{figure*}
\begin{center}
\includegraphics[width=180mm]{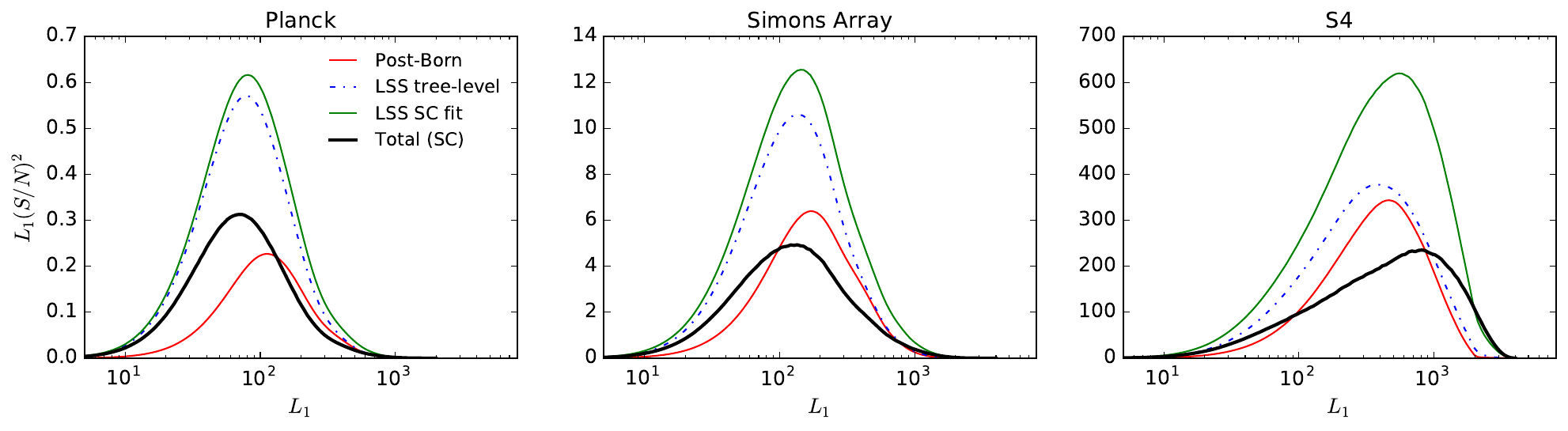}
\caption{
Contributions per log $L_1$ to the naive CMB lensing convergence bispectrum signal to noise as a function of the largest-scale mode $L_1$ for various noise levels (increase from current (Planck left), to third and fourth generation observations on the right). The red curve shows the signal from post-Born effects if it were the only contribution to the bispectrum, the green line shows the corresponding result if the only signal were from LSS (with the difference between the tree-level (blue) and fit of Ref.~\cite{Scoccimarro:2000ee} (green) giving an indication of the likely importance of fully non-tree-level effects). In reality only the total is observed, which has lower signal to noise (black line) due to partial cancellations between the different contributions.
}
\label{fig:kappaSN}
\end{center}
\end{figure*}

\begin{table}
\centering
\begin{tabular} {| c || c | c | c | c || c | c|c| c |}
\hline
 & noise $ [\muKarcmin]$ & beam $[\arcmin]$  &  $\ell_{\rm max}$ & $f_{\rm sky}$& $\Delta\kappa\kappa\, S/N$  & $\omega\omega\, S/N$ & $\kappa\kappa\kappa\, S/N$ & ${\kappa\kappa\omega}\,S/N$ \\ [0.5ex]
%heading
\hline
\hline
Planck                & 33  & 5  & 2000 & 0.7 & 0.0 & 0.0  & 0.8 & 0.1 \\
\hline
Simons Array          & 12  & 3.5& 4000 & 0.65 &0.0  & 0.0 & 3.4 & 0.4\\
\hline
SPT 3G                & 4.5 & 1.1& 4000 & 0.06  & 0.0 & 0.0 & 2.3 & 0.4 \\
\hline
S4                    & 1   &  3 & 4000  & 0.4 &0.2 & 0.7 & 25 & 3.1 \\
\hline
S5                    & 0.25&  1 & 4000  & 0.5 & 0.8 & 2.7 & 99 & 8.8 \\
\hline
\end{tabular}
%kappa bispectrum results:
%       Born Fish =  701.935, S/N =  26.494  S4_Noise1_Beam3
%     LSS_SC Fish = 1573.426, S/N =  39.666  S4_Noise1_Beam3
%LSS_SC_Born Fish =  646.611, S/N =  25.429  S4_Noise1_Beam3
%        LSS Fish =  925.929, S/N =  30.429  S4_Noise1_Beam3
%       Born Fish =    0.397, S/N =   0.630  planck
%     LSS_SC Fish =    1.185, S/N =   1.089  planck
%LSS_SC_Born Fish =    0.618, S/N =   0.786  planck
%        LSS Fish =    1.070, S/N =   1.034  planck
%       Born Fish =   12.036, S/N =   3.469  SA
%     LSS_SC Fish =   26.755, S/N =   5.173  SA
%LSS_SC_Born Fish =   11.499, S/N =   3.391  SA
%        LSS Fish =   21.652, S/N =   4.653  SA
%       Born Fish =    6.709, S/N =   2.590  SPT3G
%     LSS_SC Fish =   13.333, S/N =   3.651  SPT3G
%LSS_SC_Born Fish =    5.394, S/N =   2.322  SPT3G
%        LSS Fish =   10.257, S/N =   3.203  SPT3G
%       Born Fish = 6928.770, S/N =  83.239  Noise0.25_Beam1
%     LSS_SC Fish = 20040.287, S/N = 141.564  Noise0.25_Beam1
%LSS_SC_Born Fish = 9827.951, S/N =  99.136  Noise0.25_Beam1
%        LSS Fish = 9749.688, S/N =  98.741  Noise0.25_Beam1

\caption{
Configuration of various experiments considered for approximate signal to noise estimates assuming simplistic  lensing reconstruction noise modelling. The signal to noise ($S/N$) are given
for the post-Born correction to the convergence power spectrum ($\Delta\kappa\kappa$), the
field-rotation power spectrum from post-Born ($\omega\omega$), the
total convergence bispectrum (post-Born + LSS, $\kappa\kappa\kappa$) and the mixed convergence-rotation bispectrum (only from post-Born, $\kappa\kappa\omega$).
Multipoles used are $\ell < 4000$ in all cases\footnote{In the published version
it was intended to use only $\ell < 3000$ in TT since the very small-scale signal will be foreground dominated, but the table is for $\ell<4000$. If the $\ell<3000$ cut is made in TT the detection
significances are somewhat decreased, with $2.3 \sigma$ for $\omega\omega$ with S5, and $2.7\sigma$ for $\kappa\kappa\omega$ and $14\sigma$ for $\kappa\kappa\kappa$ with S4.
}
%In all cases we restrict the temperature to $\ell < 3000$ ,
and use
quadratic estimator lensing reconstruction noise for the minimum variance estimator~\cite{Hu:2001kj} (no iterative delensing, and taking the noise to be Gaussian and uncorrelated). The noise in the table is that for the temperature; we assume $\sqrt{2}\times$ larger noise for the polarization (except Planck, where the factor is $2\times$ as not all the detectors were polarized). The beam is given as a full-width half-maximum (FWHM). All results assume all cosmological parameters are fixed and non-linear corrections from \HALOFIT\ \cite{Takahashi:2012em} and \citet{Scoccimarro:2000ee}; post-Born effects on $C^{\kappa\kappa}_L$ are negligible, and in practice would also be swamped by non-linear modelling and parameter uncertainties.
}
\label{table:sigmas}
\end{table}

To assess the observational significance of the bispectrum we can roughly quantify the detectability by considering the optimal estimator for the bispectrum in the zero signal limit, assuming Gaussian quadratic-estimator reconstruction noise on the lensing measurement. This approximation is not expected to be accurate due no-zero signal contributions and non-Gaussian contributions to the covariance, and in principle for low noise levels it could also be possible to improve on quadratic estimators using more optimal estimators~\cite{Hirata:2003ka}. But in practice is likely to get an upper limit on how important the signal could be. We use the flat-sky Fisher matrix for the bispectrum template amplitude at zero signal~\cite{Zaldarriaga:2000ud,Hu:2000ee}
\be
 \left(\frac{S}{N}\right)^2\equiv F =\frac{f_{\rm sky}}{6\pi} \int \frac{d^2L_1d^2 L_2}{(2\pi)^2} \frac{\left(b_{L_1 L_2 L_3}^{\kappa\kappa\kappa}\right)^2}{N^\kappa_{L_1} N^\kappa_{L_2} N^\kappa_{L_3}},
\ee
where $F^{-1/2}$ gives the fractional error on the bispectrum amplitude estimate and $N^\kappa$ includes signal and noise variance on the convergence field. Fig.~\ref{fig:kappaSN} shows the contributions to the $(S/N)^2$ as a function of the largest mode for a few experimental configurations. In all cases a substantial fraction of the signal comes from semi-squeezed and folded shapes involving at least one relatively large scale mode\footnote{For S4 and Planck, just over half the $(S/N)^2$ is from $L_2,L_3 > 2L_1$; for S4 for the signal from $L_2,L_3>4L_1$ would be detectable at $\sim 5\sigma$.}, and the signal to noise including both post-Born and LSS contributions is significantly lower than if either had been present separately due to the partial cancellation. Beyond tree-level contributions to the bispectrum (`LSS SC fit') give a noticeable but modest increase in signal, but could be somewhat larger with a more realistic fully non-linear LSS bispectrum calculation.

The possible signal to noise for various observational configurations is shown in Table.~\ref{table:sigmas}.
In the case of Planck-like noise, the LSS signal alone would ideally be (un)detectable at about $1.1\sigma$, which the addition of the post-Born contributions lowers to $0.8\sigma$. For more sensitive observations the signal is much more important, and would potentially be detectable at $\sim 25\sigma$ with `S4' observations with $1\muKarcmin$ noise and few-arcminute beam. At low noise levels, where smaller scales modes become resolved and contribute to the largest-scale in the bispectrum signal, the reduction in signal from post-Born contributions remains significant: without the post-Born contributions the $\sim 25\sigma$ would have been $\sim 40\sigma$: post-Born contributions  are important to correctly model the shape and amplitude of the bispectrum signal.

 The impact of a lensing convergence bispectrum on lensing power spectrum estimation is investigated in detail in Ref.~\cite{Bohm:2016gzt} and shown to lead to a non-negligible bias on some lensing reconstruction quadratic estimators for low noise levels. These biases be thought of as particular (non-optimal) estimators for the bispectrum amplitude. As such they can be expected to have substantially lower significance than the optimal estimator, for example for Planck noise levels biases due to the bispectrum should be negligible (much less than $0.8\sigma$, in agreement with Ref.~\cite{Bohm:2016gzt}).

\subsection{The $\kappa\kappa\omega$ bispectrum}

\begin{figure*}
\begin{center}
\includegraphics[width=180mm]{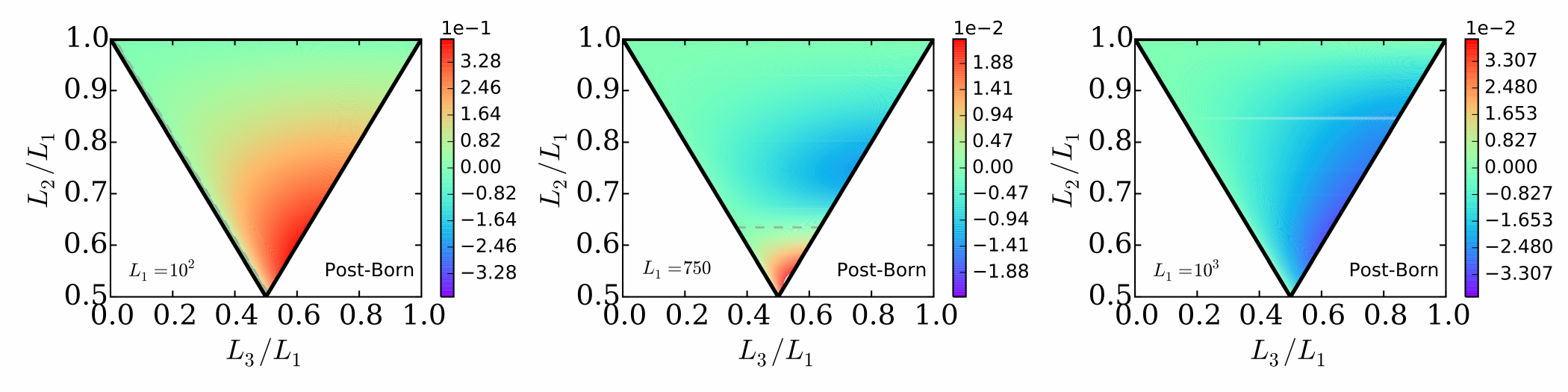}
\caption{
{
Slices through the weighted shear-rotation bispectrum $(L_2 L_3)^{1/2} \, b^{\kappa\kappa\omega+}_{L_1 L_2 L_3} / (C^{\kappa\kappa}_{L_1} C^{\kappa\kappa}_{L_2} C^{\omega\omega}_{L_3})$ for $L_1 = 10^2$ (left plot), $750$ (middle plot) and $10^3$ (right plot). The grey dashed line denotes the $b^{\kappa\kappa\omega}=0$ contour.
}
}
\label{fig:bispectrum_omegakk_PB_LSS_Shape}
\end{center}
\end{figure*}

The field rotation is a distinctive signal of lens-lens coupling and strongly non-Gaussian, giving a distinctive bispectrum.
%Since the field rotation signal is very small, we only consider bispectra involving one rotation mode coupled to two convergence modes.
Working to $\mathcal{O}(\Psi^4)$, we only need to consider the bispectra involving one rotation mode coupled to two convergence modes, $\langle \kappa \kappa \omega \rangle$, as other bispectra combinations, such as $\langle \kappa \omega \omega \rangle$ or $\langle \omega \omega \omega \rangle$, only appear at higher orders.
Eq.~\eqref{eq:kkz} then gives the mixed bispectrum
\be
\label{eqn:kkw_PB}
b_{L_1L_2L_3}^{\kappa\kappa\omega\pm} = -\sin 2\varphi_{21}\left[ \Mkappa_s(L_1,L_2)-\Mkappa_s(L_2,L_1)\right],
\ee
where $\pm$ denotes the sign of $\sin \varphi_{21}$ (the bispectrum has odd parity, so the lengths of the vectors do not uniquely define the sign).
Numerical slices through the bispectrum are shown in Figs.~\ref{fig:bispectrum_omegakk_PB_LSS_Shape}.
In this case the naive Fisher signal-to-noise is given by
\be
 \left(\frac{S}{N}\right)^2\equiv F =\frac{f_{\rm sky}}{2\pi} \int \frac{d^2L_1d^2 L_2}{(2\pi)^2} \frac{\left(b_{L_1 L_2 L_3}^{\kappa\kappa\omega}\right)^2}{N^\kappa_{L_1} N^\kappa_{L_2} N^\omega_{L_3}}
\ee
assuming the lensing reconstruction noise on the rotation and convergence are uncorrelated. Although the field rotation is very small, there is no linear signal, so there is no leading-order cosmic variance contribution to the rotation noise $N^\omega_{L_3}$,
which makes the mixed bispectrum potentially detectable with future observations. The lensing reconstruction noise on the $\omega$ is also lower than that on $\kappa$ on large scales when using $EB$ reconstruction estimators.

Table~\ref{table:sigmas} shows that with idealized assumptions the mixed bispectrum could become detectable with stage four (S4) observations and beyond, but is negligible for observations currently underway.  The rotation bispectrum could also source a bias in deflection quadratic estimators, similar to the effect from the LSS bispectrum discussed in Ref.~\cite{Bohm:2016gzt}. Since the optimal estimator for the bispectrum is currently negligible, the bias should also be negligible, however in future once the bispectrum becomes a substantial signal the rotation contribution to the bias should also be accounted for.

\section{CMB B modes and CMB power spectra}
\label{sec:CMBpower}

\begin{figure*}
\begin{center}
\includegraphics[width=140mm]{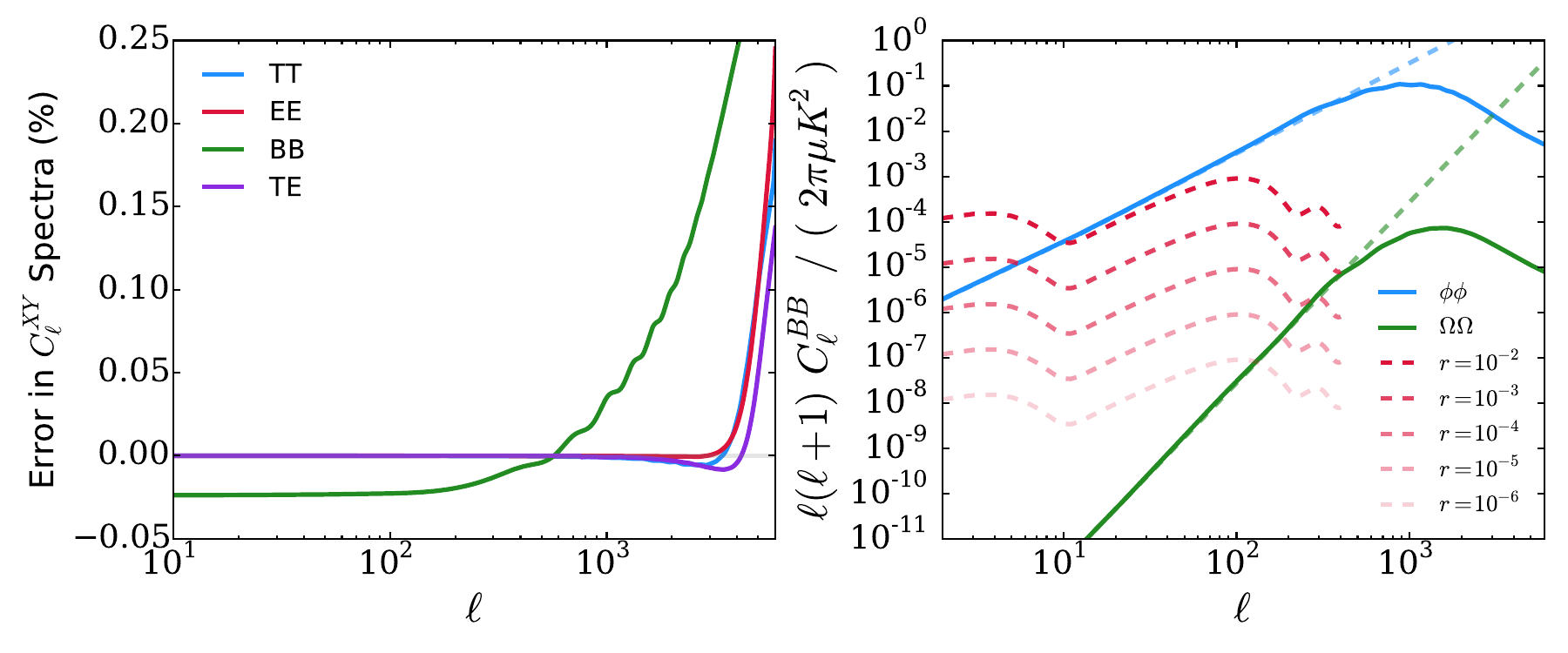}
\caption{
%\GP
{The left plot shows the percentage change on the lensed CMB power spectra $C^{XY}_{\ell}$, with $X,Y \in \lbrace T,E,B \rbrace$, induced by post-Born corrections. We have taken the total lensing contribution sourced by the both the potential $C^{\phi \phi}_{\ell}$ and the curl $C^{\Omega\Omega}_{\ell}$ components. For $\ell \lesssim 3000$, the post-Born corrections to the TT, EE and TE spectra are $\lesssim$ 0.05\% and therefore negligible. The lensed $C_{\ell}$ were calculated using the leading-order perturbative flat-sky result. The right plot shows BB power spectra induced by lensing of scalar E-mode polarization, separately for the potential $\phi$ (blue, convergence) and the curl $\Omega$ (green, rotation) contributions. The dashed lines show the low-$\ell$ approximations as given by Eq.~\eqref{Eqn:BB_lowl}. The red-dashed lines show the BB power spectra sourced by primordial gravitational waves with various values of the tensor-to-scalar ratio $r$. For $\ell \sim 100$, the post-Born curl contributions to the BB spectrum can be comparable to the contribution from primordial gravitational waves for $r \lesssim 10^{-6}$.
All results shown here neglect contributions from non-Gaussianity of the lensing field.
}
}
\label{fig:Error_Polarisation_BB}
\end{center}
\end{figure*}

Lensing affects the CMB power spectra by smoothing out the peaks, transferring power to small scales, and producing B-mode polarization. The corresponding lensed power spectra can be calculated using standard methods~\cite{Seljak:1996ve,Challinor:2005jy}. Since the post-Born contributions to the lensing power spectra are much smaller than the leading contribution,
for temperature and E-mode polarization the change in lensed spectra due to the additional post-Born contributions is negligible: the tiny $\alt 0.2\%$ changes are shown in Fig.~\ref{fig:Error_Polarisation_BB}.
This is however assuming the lensing potential is Gaussian, which is not entirely accurate since post-Born and LSS growth both induce a non-zero bispectrum. Contributions from the post-Born convergence bispectrum are somewhat larger than the tiny effect shown in Fig.~\ref{fig:Error_Polarisation_BB} from the change in the lensing power spectrum~\cite{Marozzi:2016uob}. However, the convergence bispectrum actually has larger contributions from the LSS bispectrum, so a full calculation must include both. Since it is not predominantly a post-Born effect, which is the focus of this paper, we refer the reader to detailed calculations in Ref.~\cite{Lewis:2016tuj}. There we show that the total bispectrum contribution still only changes the lensed CMB power spectra at a level that negligible for the foreseeable future.

%\GP{Shortly after this paper was submitted, a calculation of the leading effect of the post-Born bispectrum on the lensed CMB temperature spectra was presented in Ref.~\cite{Marozzi:2016uob}, but as they did not include the LSS contribution their result is not directly observable. In addition, Ref.~\cite{Marozzi:2016uob} misestimates the sign of the full bispectrum contribution as the LSS terms will have opposite sign and will be some what larger with enhancements to the total bispectrum only appearing in the equilateral limit, which is subdominant to non-equilateral configurations in CMB weak lensing \cite{Pratten:2016temp}.}

For the B modes the post-Born effects are potentially more important, since on small scales, B-mode polarization from primordial gravitational waves is expected to be negligible and the signal is almost entirely from the B modes from lensing. The dominant signal is from lensing by the linear CMB lensing potential, and the corresponding lensed B-mode power spectrum is easily calculated accurately following Ref.~\cite{Challinor:2005jy}. Here, we focus on the small additional contributions from post-Born lensing, and use the leading flat-sky series-expansion approximation which captures the main effect accurately: the lensed polarization tensor $\tilde{P}_{ab}$ can be expanded in terms of the unlensed tensor $P_{ab}$ as\footnote{We neglect the second order effect of polarization rotation due to parallel transport which is  $\clo(\alpha^2)$ (and hence much smaller than $\clo(\kappa^2)$ effects like image rotation). The negligible corrections from polarization rotation (and additional larger, but still small corrections due to emission angle and time delay) are calculated in Ref.~\cite{Lewis:2017ans}.}
\be
\tilde{P}_{ab}(\vtheta) = P_{ab}(\vtheta + \valpha) \approx
P_{ab}(\vtheta)+ \valpha_c\grad^c P_{ab}(\vtheta) + \frac{1}{2} \valpha_c\valpha_d\grad^c\grad^d P_{ab}(\vtheta).
\ee
Assuming negligible unlensed B modes, using Eq.~\eqref{eq:alphadecomp} to decompose the deflection in terms of curl and gradient components, we get the Fourier-space expansion~\cite{Hu:2000ee,Lewis:2006fu}
\begin{multline}
%Feb17 fixed sign
\tilde{B}(\vell) = - \int \frac{d^2\vell'}{(2\pi)^2} E(\vell') \sin (2 \varphi_{\vell'\vell}) \left[
\vell'\times \vell \Omega(\vell-\vell') + (\vell-\vell')\cdot \vell' \phi(\vell-\vell')
\right]
\\
-\frac{1}{2} \int \frac{d^2\vell_1}{(2\pi)^2}\int \frac{d^2\vell_2}{(2\pi)^2} \sin(2\varphi_{\vell_1\vell})
E(\vell_1)
\left[\vell_1\cdot \vell_2 \phi(\vell_2) + \vell_1\times\vell_2\Omega(\vell_2)\right]\times\\
\left[ \vell_1\cdot (\vell_1+\vell_2 - \vell)\phi(\vell-\vell_1-\vell_2)  + \vell_1\times(\vell_2-\vell)
\Omega(\vell-\vell_1-\vell_2) \right],
\label{eq:Bexfull}
\end{multline}
where $\varphi_{\vell'\vell}$ is the angle between $\vell'$ and $\vell$ and $E(\vell)$ are the unlensed E-modes.
The dominant contribution comes from the first linear lensing potential ($\phi$) term.
The small post-Born correction to the gradient potential $\phi$ only has a small fractional effect, but the curl contribution ($\Omega)$
is potentially more interesting: it has different symmetry properties, and for a given fixed spectrum shape rotations generate small-scale B modes more efficiently than gradient deflections~\cite{Padmanabhan:2013xfa}.
Note that since $\Omega$ from post-Born is second order,
$\phi$ contributions to the second term are the same order in $\Psi$ as the (small) leading $\Omega$ contribution.

We first focus on the leading rotation-induced B modes given by~\cite{Hirata:2003ka,Cooray:2005hm}
\be
%Feb17 fixed sign
\tilde{B}(\vell, \text{rotation}) =  \int \frac{d^2\vell'}{(2\pi)^2} E(\vell') \sin (2 \varphi_{\vell'\vell})
\,\,\vell\times \vell' \,\Omega(\vell-\vell').
\label{eq:Bexand}
\ee
The leading purely rotation-sourced B-mode power spectrum is then\footnote{
This result differs from Eq. 52 in Ref.~\cite{Hirata:2003ka}, which uses an incorrect definition of the angle $\alpha$ in their $\sin^2(2\alpha)$ factor. Our numerical result is significantly smaller at $\ell \sim 1000$, but comparable on large scales.
}
\be
\tilde{C}_\ell^{BB}({\rm rotation}) = 4\int \frac{d^2 \bfell'}{(2\pi)^2} \sin^2 (\varphi_{\vell'\vell}) \sin^2 (2 \varphi_{\vell'\vell}) \frac{\ell^2{\ell'}^2}{{\ell''}^4}C^{\omega\omega}_{\ell''} C^{EE}_{\ell'},
\label{eq:BBrot}
\ee
where $C_\ell^{EE}$ is the unlensed E-mode polarization power spectrum and $\vell''=\vell-\vell'$. The numerical result is shown in Fig.~\ref{fig:Error_Polarisation_BB}.
For low $\ell$, Eq.~\eqref{eq:BBrot} is approximately given by
\be
\tilde{C}_{\ell}^{BB}({\rm rotation}) \approx \frac{\ell^2}{2\pi} \int d \ln \ell' C^{\omega\omega}_{\ell'} C^{EE}_{\ell'}.
\ee
%\be
%\tilde{C}_{\ell}^{BB}({\rm convergence}) \approx \frac{1}{\pi} \int d \ln \ell' C^{\kappa\kappa}_{\ell'} \ell'{}^2C^{EE}_{\ell'}.
%\ee
On large scales it therefore scales as $\ell^2$, and has a steeper slope than the gradient B-mode which has a white-noise spectrum. Specifically, at large scales we have
\be
\label{Eqn:BB_lowl}
\tilde{C}_\ell^{BB}({\rm{convergence}}) \approx 2.0\times 10^{-6}\muK^2,
\qquad
\tilde{C}_\ell^{BB}({\rm{rotation}}) \approx 1.7\times 10^{-11} \left(\frac{\ell}{100}\right)^2 \muK^2.
\ee
At $\ell\sim 100$ most relevant for primordial gravitational waves the lensing B-mode power sourced by field rotation is five orders of magnitude lower than the dominant convergence signal: it will not be an important source of confusion for gravitational waves with $r\agt 10^{-6}$ (see Fig.~\ref{fig:Error_Polarisation_BB}).
The qualitatively different large-scale behaviour appears because the leading-order curl contribution is a total derivative: the leading-order lensed polarization tensor is $\valpha_c\nabla^c P_{ab}$, where for a curl deflection $\valpha_c\nabla^c P_{ab} = \epsilon_{cd}\nabla^d\Omega \nabla^c P_{ab}=\epsilon_{cd}\nabla^d(\Omega \nabla^c P_{ab})$. Small-scale modes of $\Omega$ and $P_{ab}$ give, on large scales, an uncorrelated contribution to $\Omega \nabla^c P_{ab}$, which therefore has a white spectrum like the gradient-induced B modes. But only the total derivative enters the lensed polarization, so the curl component has a blue spectrum and is suppressed on large scales.

On smaller scales the B-mode rotation becomes relatively more important, and the geometrical enhancement of the lensed B-mode production described by Ref.~\cite{Padmanabhan:2013xfa}, makes the total fractional effect on the BB power spectrum fractionally somewhat larger than the fractional effect from the gradient lensing spectrum. The effect on the BB power spectrum is still small though, $\alt 0.2\%$ at $\ell < 3000$. At the map level the rotation B-mode contribution makes about 2.5\% of the total lensed B-mode r.m.s., but only affects power spectra at a low level.

\begin{figure*}
\begin{center}
\includegraphics[width=80mm]{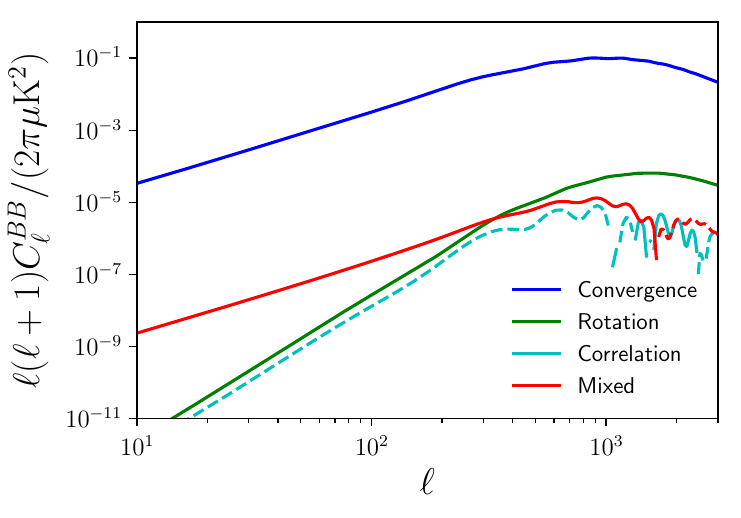}
\caption{
Contributions to the lensed BB polarization power spectrum involving post-Born rotation. The green line shows the result of Eq.~\eqref{eq:Bexand} from the leading rotation term squared, and for reference the blue line shows the standard lensing B-mode power spectrum from convergence. The other two lines show couplings between rotation and convergence proportional to the bispectrum: the cyan line shows the $\clo(\Omega)\times \clo(\phi^2)$ contribution from the correlation between the leading rotation and the second order convergence contributions (Eq.~\ref{eq:BBcorrelation}); red shows the $\clo(\phi\Omega)\times \clo(\phi)$ term from correlating the leading convergence contribution with the mixed contribution to the second order term (Eq.~\ref{eq:BBmixed}).
On large scales the mixed term dominates the rotation contributions as it has a white rather than blue spectrum; both bispectrum terms are very small on small scales. Dashed lines denote negative contributions.
}
\label{fig:BBbi}
\end{center}
\end{figure*}

The rotation modes of Eq.~\eqref{eq:Bexand} are uncorrelated to the dominant leading-order convergence modes, since $C_L^{\kappa\omega}=0$. However there are additional couplings from the $b^{\phi\phi\Omega}$ bispectrum that lead to a small correlation. Specifically, contributions to the BB power spectrum from a coupling between  Eq.~\eqref{eq:Bexand} and the second-order term involving the product of two $\phi$s in Eq.~\eqref{eq:Bexfull} is
\begin{multline}
%Feb17 fixed sign
\tilde{C}^{BB}_{\ell}(\text{correlation}) = \int \frac{d^2 \vell_1}{(2 \pi)^2} \int \frac{d^2 \vell_2}{(2 \pi)^2} \, \left[ \vell_1 \cdot \vell_2 \right] \left[ \, \bfell \times \vell_1 \right] \, \left[ \vell_1 \cdot (\vell_2 + \vell_1 + \bfell ) \right] \, \sin^2 (2 \varphi_{\vell_1 \bfell}) \\
   \times C^{EE}_{\ell_1} \, b^{\phi\phi\Omega} (\vell_2 , -\vell_1 - \vell_2 - \bfell,\bfell + \vell_1) .
 \label{eq:BBcorrelation}
\end{multline}
The post-Born bispectrum is given by Eq.~\eqref{eqn:kkw_PB} and the numerical result for the corresponding contribution to BB is shown in Fig.~\ref{fig:BBbi}.
 Since it only comes from a correlation involving the sub-leading convergence modes, it is the same order as
$\tilde{C}_\ell^{BB}({\rm{rotation}})$, but substantially smaller on small scales. The leading rotation-induced B contributions are therefore well approximated as uncorrelated to the convergence-induced B modes.
The leading effect on lensing convergence reconstruction estimators should therefore be similar to uncorrelated noise component, and should not substantially limit gradient lensing reconstruction for the near future~\cite{Hirata:2003ka}. In principle the curl component can also be reconstructed and partially delensed as for the dominant gradient component: maximum-likelihood reconstruction can just estimate the full $\valpha$ field.

There is a further  contribution to the lensed BB power spectrum involving the rotation, a $\clo(\phi)\times \clo(\phi\Omega)$ term from the correlation of the leading convergence mode with the mixed convergence-rotation contribution to the second order term. It also involving the bispectrum and is the same order in $\Psi$:
\begin{multline}
\tilde{C}^{BB}_{\ell}(\text{mixed}) = -2\int \frac{d^2 \vell_1}{(2 \pi)^2} \int \frac{d^2 \vell_2}{(2 \pi)^2} \,
\left[ \vell_1 \cdot  \vell_2 \right] \,
\left[ \vell_1 \cdot (\vell+\vell_1) \right]
\left[ \vell_1 \times (\vell_2 + \vell) \right] \, \sin^2 (2 \varphi_{\vell_1 \bfell}) \\
   \times C^{EE}_{\ell_1} \, b^{\phi\phi\Omega} (\vell+\vell_1 , \vell_2,-\vell-\vell_1-\vell_2).
 \label{eq:BBmixed}
\end{multline}
This term is shown in Fig.~\ref{fig:BBbi}, and in fact dominates the rotation contributions on large scales because it has a white rather than blue spectrum there. On small scales it is much smaller than the leading rotation contribution. Although Eq.~\eqref{eq:BBmixed} dominates the curl contributions to the power, it should not in itself substantially impact residual delensed B-mode power: delensing substantially reduces the leading $\phi$ contribution at the map level. Compared to the dominant convergence B-mode power without delensing, it is four orders of magnitude smaller and hence negligible.

As mentioned previously, there are also additional terms involving the convergence bispectrum that contribute to all the lensed CMB power spectra; here we have focussed only on  the distinctively post-Born rotation, see Ref.~\cite{Lewis:2016tuj} for full details of the other contributions.

\section{Conclusions}

Post-Born lensing of the CMB has several small but interesting effects, some of which will become observationally relevant quite soon. Specifically, post-Born lensing:

\begin{itemize}
  \item Gives corrections to the lensing convergence power spectrum, but only at the $0.2\%$ level, which is small enough to be negligible (in agreement with Refs.~\cite{Krause:2009yr,Calabrese:2014gla}, but contrary to Ref.~\cite{Hagstotz:2014qea}).
  \item Introduces a field rotation (curl) power spectrum, which will remain negligible for the near future, but ultimately may be detectable using curl lensing reconstruction~\cite{Hirata:2003ka}.
  \item Contributes significantly to the convergence bispectrum, at a comparable amplitude to the signal from non-linear large-scale structure growth. For flat (mode-aligned) shapes with largest post-Born signal, there is a partial cancellation with the LSS bispectrum, reducing the total bispectrum signal (but remaining detectable with upcoming observations). Including the post-Born convergence bispectrum signal will be important to get the correct shape and amplitude of the full signal, and hence cosmology from the bispectrum~\cite{Namikawa:2016jff} and bias on quadratic estimators~\cite{Bohm:2016gzt}.
  \item Produces a mixed convergence-rotation bispectrum, which may be detectable for observations with $\alt \muKarcmin$ noise and few-arcminute beam.
  \item Modifies the lensed temperature and E-polarization CMB power spectra, but only at a negligible $\alt 0.2\%$ level.  The full leading effect of post-Born and LSS lensing non-Gaussianities on CMB power spectra is calculated in Ref.~\cite{Lewis:2016tuj}, and does not change this conclusion.
  \item Produces a distinctive lensed B-mode from field rotation, which to leading order is a total derivative and hence has a blue spectrum on large scales. The signal is negligible compared to gravitational waves with $r\agt 10^{-5}$. On small scales rotation B modes contributes about $2.5\%$ of the B-mode amplitude, but are largely uncorrelated to the larger gradient contribution~\cite{Hirata:2003ka}.
   \item Produces contributions to the lensed CMB power spectra from  non-Gaussianity of the lensing field which slightly correlate the rotation and convergence B modes.
       On large scales the dominant curl-dependent contribution to the lensed BB power is from a mixed bispectrum coupling with a white spectrum (but still four orders of magnitude smaller than the dominant convergence B-mode power).
\end{itemize}

We have only considered bispectra involving the lensing convergence or rotation in detail, but post-Born corrections are also expected to be relevant for bispectra involving CMB lensing and large-scale structure (and galaxy lensing). Mixed bispectra may provide a probe of the rotation signal, and the post-Born and LSS convergence bispectrum could lead to a bias to the cross-correlation between the reconstructed lensing potential and tracers of large-scale structure. Future work should investigate these in more detail, along with a detailed calculation of the bias on lensing reconstruction power spectrum estimators using the full LSS plus post-Born bispectrum~\cite{Bohm:2016gzt}.

There still remains uncertainty in the detailed modelling of the non-linear matter distribution. In this work we adopted variations of the \HALOFIT\ non-linear matter power spectrum, and simple fits for the matter bispectrum, both calibrated to N-body simulations. Uncertainty in these predictions are far more important than post-Born contributions for the lensing power spectrum, and limit the accuracy to which the total bispectrum can be calculated.

%%%%%%%%%%%%%%%%%%%%%%%%% Acknowledgements %%%%%%%%%%%%%%%%%%%%%%%%%
\section*{Acknowledgments}
We thank Vanessa B\"ohm, Anthony Challinor, Giulio Fabbian, Marcel Schmittfull and Blake Sherwin for comments, discussion and comparison of some numerical results.
We acknowledge support from the European Research Council under
the European Union's Seventh Framework Programme (FP/2007-2013) / ERC Grant Agreement No. [616170]. The results in this paper were derived using Julia \cite{Bezanson:2012,Bezanson:2014} and Python.
Part of this work was undertaken on the COSMOS Shared Memory system at DAMTP, University of Cambridge operated on behalf of the STFC DiRAC HPC Facility. This equipment is funded by BIS National E-infrastructure capital grant ST/J005673/1 and STFC grants ST/H008586/1, ST/K00333X/1.

%%%%%%%%%%%%%%%%%%%%%%%% Appendices %%%%%%%%%%%%%%%%%%%%%%%%%
\appendix

\section{Post-Born source terms}
\label{app:PBA}
\subsection{First Order}
At first order, we have just the one trivial source term:
\begin{align}
\label{eqn:S1}
S^{(1)}_{ab} (\chi) &= \Psi_{,ab} (\chi) .
\end{align}

\subsection{Second Order}
At second order, we have the following source terms:
\begin{align}
\label{eqn:S2L}
S^{(2L)}_{ab} (\chi) &= -2 \int^{\chi}_{0} d \chi^{\prime} \, \chi^{\prime 2} \, W ( \chi^{\prime} , \chi ) \, \left[  \Psi_{,ac} (\chi) \Psi_{,cb} (\chi^{\prime}) \right] , \\
S^{(2D)}_{ab} (\chi) &= -2 \int^{\chi}_0 d \chi^{\prime} \, \chi^{\prime} \chi \, \, W ( \chi^{\prime} , \chi ) \, \left[ \Psi_{,abc} (\chi) \Psi_{,c} (\chi^{\prime}) \right] .
\label{eqn:S2B}
\end{align}

\subsection{Third Order}
\n
At third order, we have the following source terms\footnote{Note that $\psi^{(3D)}_{ab} = \psi^{(3A)}_{ab}$ in \cite{Krause:2009yr}, $\psi^{(3X)}_{ab} = \psi^{(3B)}_{ab}$ in \cite{Krause:2009yr} and $\psi^{(3L + 3X1 + 3X2)}_{ab} = \psi^{(3C)}_{ab}$ in \cite{Krause:2009yr}. }:
\begin{align}
S^{(3D)}_{ab} (\chi) &= +2 \chi^2 \int^{\chi}_0 d \chi^{\prime} \, W (\chi^{\prime} , \chi ) \, \chi^{\prime} \, \int^{\chi}_0 d \chi^{\prime \prime} W(\chi^{\prime \prime} , \chi) \, \chi^{\prime \prime} \, \left[ \Psi_{,abcd} (\chi) \Psi_{,c} (\chi^{\prime}) \Psi_{,d} (\chipp) \right] ,
\label{eqn:S3B} \\
S^{(3X)}_{ab} (\chi) &= +4 \chi \int^{\chi}_0 d \chi^{\prime} W (\chi^{\prime} , \chi ) \, \chi^{\prime 2} \, \int^{\chi}_0 d \chi^{\prime \prime} W(\chi^{\prime \prime} , \chi) \, \chi^{\prime\prime} \, \left[ \Psi_{,acd} (\chi) \Psi_{,bc} (\chi^{\prime}) \Psi_{,d} (\chi^{\prime \prime}) \right] ,
\label{eqn:S3X} \\
S^{(3L)}_{ab} (\chi) &= +4 \int^{\chi}_0 d \chi^{\prime} W (\chi^{\prime} , \chi ) \, \chi^{\prime 2} \, \int^{\chi^{\prime}}_0 d \chi^{\prime \prime} W(\chipp , \chip) \, \chi^{\prime \prime 2} \, \left[ \Psi_{,cd} (\chi) \Psi_{,cb} (\chip) \Psi_{,bd} (\chipp) \right] ,
\label{eqn:S3L} \\
S^{(3X1)}_{ab} (\chi) &= +4 \chi \int^{\chi}_0 d \chi^{\prime} W (\chi^{\prime} , \chi ) \, \chi^{\prime 2} \, \int^{\chi^{\prime}}_0 d \chi^{\prime \prime} W(\chipp , \chip) \, \chipp \, \left[ \Psi_{,d} (\chipp) \Psi_{,cd} (\chip) \Psi_{,abc} (\chi) \right] ,
\label{eqn:S3X1} \\
S^{(3X2)}_{ab} (\chi) &= +4 \int^{\chi}_0 d \chi^{\prime} W(\chi^{\prime} , \chi ) \, \chi^{\prime 3} \, \int^{\chi^{\prime}}_0 d \chi^{\prime \prime} W(\chipp , \chip) \, \chipp \, \left[ \Psi_{,d} (\chipp) \Psi_{,ac} (\chi) \Psi_{,bcd} (\chip) \right] .
\label{eqn:S3X2}
\end{align}

\section{Derivation of power spectrum corrections}
\label{app:PBC}

In this appendix we explicitly derive the post-Born corrections at order $\mathcal{O}(\Psi^4)$. Under the Limber approximation, the only surviving ``13" term is given by a 13D ray-deflection correction to the convergence and shear. The ``22" terms reduce to a particularly convenient expression with only the 22L ``lens-lens" term contributing to the curl spectrum. The results in this appendix manifestly agree with Ref. \cite{Krause:2009yr} and Ref. \cite{Cooray:2002mj}. We also explicitly show that the terms in Ref. \cite{Shapiro:2006em} vanish as noted by Ref. \cite{Krause:2009yr}.

\subsubsection{22D Correction}
The 22D correction corresponds to second-second order ray deflections\footnote{Note that in Ref. \cite{Cooray:2002mj} the ray-deflection corrections are called ``Born corrections" and denoted with ``B". The the lens-lens and cross terms are labelled ``L" and ``X" respectively.}. Schematically, the structure of the 22D correction is given by
\begin{multline}
\label{eqn:22DFull}
\langle \psi^{(22D)}_{ab}  (\bfell,\chi_s) \, \psi^{(22D)}_{cd} (\bfellp , \chi_s) \rangle = \int^{\chi_s}_0 d\chi \, \int^{\chi_s}_0 d \chip \, \int^{\chip}_0 d \chipp \int^{\chip}_0 d \chippp \, \chi^2 \, W(\chi,\chi_s) \, \chi^{\prime 2} \, W(\chip,\chi_s) \, \chi^{\prime \prime} W(\chipp,\chip) W(\chippp,\chip) \\
\times (2 \pi)^4 \int \frac{d^2 \bfL}{\dlw} L_a L_b L_g \int \frac{d^2 \bfLp}{\dlw} L^{\prime}_c L^{\prime}_d L^{\prime}_f \int \frac{d^2 \bfLpp}{\dlw} L^{\prime \prime}_g  \int \frac{d^2 \bfLppp}{\dlw} L^{\prime \prime \prime}_f  \,\delta_D (\bfL + \bfLp - \bfell) \delta_D (\bfellp + \bfLp + \bfLppp) \\
\times \frac{1}{\chi^3 \chipp \chip^3 \chippp} \left\langle {\Psi} (\chi,\bfL) {\Psi} (\chip,\bfLp) {\Psi} (\chipp,\bfLpp) {\Psi} (\chippp,\bfLppp) \right\rangle + \lbrace ab \leftrightarrow cd \rbrace .
\end{multline}
\n
However, a number of cancellations occur upon evaluating the Wick contractions for the angular Fourier transform of the potentials ${\Psi}$. Notably, we will find two vanishing and one non-vanishing contribution. The first vanishing Wick contraction is given by
\begin{align}
\langle
\contraction{}{{\Psi}}{{}_1(\mathbf{q}_1)\>\>\>\>{}}{{\Psi}}
  \nomathglue{%
     {\Psi} (\chi,\bfL)\>\>{}{\Psi} (\chipp, \bfLpp)}
\bcontraction{}{{\Psi}}{{}(\chipp,\bfLpp)\>{}}{{\Psi}} \;
  \nomathglue{%
     {\Psi}(\chip , \bfLp)\>\>{}{\Psi} (\chippp , \bfLppp)}
\rangle
\propto \delta_D (\chi - \chipp) \, \delta_D (\chip - \chippp) \, \delta_D (\bfL + \bfLpp) \, \delta_D (\bfLp + \bfLppp) ,
\end{align}
\n
where we have used Eq.~(\ref{eqn:LimberPS}) to evaluate the contractions. This term collapses, however, as $\delta_D (\chip - \chippp) W(\chippp,\chip) = W(\chip,\chi) = 0$. The second vanishing Wick contraction is given by
\begin{align}
\langle
\contraction{}{{\Psi}}{{}_1(\mathbf{q}_1)\>\>\>\>{}}{{\Psi}}
  \nomathglue{%
     {\Psi} (\chi,\bfL)\>\>{}{\Psi} (\chippp, \bfLppp)}
\bcontraction{}{{\Psi}}{{}(\chipp,\bfLpp)\>{}}{{\Psi}} \;
  \nomathglue{%
     {\Psi}(\chip , \bfLp)\>\>{}{\Psi} (\chipp , \bfLpp)}
\rangle
\propto \delta_D (\chi - \chippp) \, \delta_D (\chip - \chipp) \, \delta_D (\bfL + \bfLppp) \, \delta_D (\bfLp + \bfLpp) ,
\end{align}
\n
which also vanishes as $\delta_D (\chip - \chipp) \delta_D (\chi - \chippp) W(\chippp,\chip) W(\chipp,\chi) = W(\chi,\chip) W(\chip,\chi) = 0$. The only non-vanishing Wick contraction is given by
\begin{align}
\langle
\contraction{}{{\Psi}}{{}_1(\mathbf{q}_1)\>\>\>\>{}}{{\Psi}}
  \nomathglue{%
     {\Psi} (\chi,\bfL)\>\>{}{\Psi} (\chip, \bfLp)}
\bcontraction{}{{\Psi}}{{}(\chipp,\bfLpp)\>{}}{{\Psi}} \;
  \nomathglue{%
     {\Psi}(\chipp , \bfLpp)\>{}{\Psi} (\chippp , \bfLppp)}
\rangle
\propto \delta_D (\chi - \chip) \, \delta_D (\chipp - \chippp) \, \delta_D (\bfL + \bfLp) \, \delta_D (\bfLpp + \bfLppp) ,
\end{align}
\n
which produces a term $W^2 (\chi,\chi_s) W^2 (\chipp,\chi)$. This final contribution leads to $\bfLp \rightarrow -\bfL$ and $\bfLppp \rightarrow -\bfLpp$ which leads to a kernel $\propto \bfL^4 ( \bfL \cdot \bfLpp)^2$. When coupled with the delta functions on the second line of Eq.~(\ref{eqn:22DFull}), we will find a term $\propto \bfL^4 \, ( \bfL \cdot (\bfell - \bfL) )^2$ where the integration is carried out over $\bfL$.

The non-vanishing contributions will therefore reduce to the following \cite{Cooray:2002mj,Krause:2009yr}
\begin{align}
\Delta C^{22D}_{abcd} (\ell) &= 16 \int \frac{d^2 \bfellp}{(2 \pi)^2} \, \ell^{\prime}_a \ell^{\prime}_b \ell^{\prime}_c \ell^{\prime}_d \left( \bfellp \cdot \bfellpp \right)^2 \, M (\ellp , \ell^{\prime \prime}) .
\end{align}
\n
where $\bfellpp = \bfell - \bfellp$, we have introduced a relabelling $\bfL = \bfell^{\prime}$ and $M(\ell^{\prime},\ell^{\prime \prime})$ is the mode coupling kernel as in Eq. \ref{eqn:MKH}. The 22D correction is expressible via an integration over a common kernel for the various lensing observables \cite{Cooray:2002mj}
\begin{align}
\label{eqn:22D}
\Delta C^{\mu \nu, 22D}_{\ell} &= 4 \int \frac{d^2 \bfellp }{(2 \pi)^2} \, \ell^{\prime 4} G^{(22D)}_{\mu} G^{(22D)}_{\nu} \left[ \bfellp \cdot \bfellpp \right]^2 M(\ell^{\prime} , \ell^{\prime \prime}) ,
\end{align}
\n
where, following \cite{Cooray:2002mj}, we have introduced a set of geometric factors for the weak-lensing observables
\begin{align}
G^{(22D)}_{\kappa} = 1, \quad G^{(22D)}_{\gamma_E} = \cos (2 \, \varphi_{\ell^{\prime}}), \quad G^{(22D)}_{\gamma_B} = \sin ( 2 \, \varphi_{\ell^{\prime}}) , \quad G^{(22D)}_{\omega} = 0.
\end{align}

\subsubsection{22L Correction}
The other second-second correction, the 22L correction, also provides a non-vanishing contribution to the post-Born power spectrum. Much as we did with the 22D correction, the 22L corrections are schematically given by
\begin{multline}
\label{eqn:22LFull}
\langle \psi^{(22L)}_{ab}  (\bfell,\chi_s) \, \psi^{(22L)}_{cd} (\bfellp , \chi_s) \rangle = \int^{\chi_s}_0 d\chi \, \int^{\chi_s}_0 d \chip \, \int^{\chip}_0 d \chipp \int^{\chip}_0 d \chippp \, \chi^3 W(\chi,\chi_s) \chi^{\prime 3} \, W(\chip,\chi_s) \chipp W(\chipp,\chip) \chippp W(\chippp,\chip) \\
\times (2 \pi)^4 \int \frac{d^2 \bfL}{\dlw} L_a L_g  \int \frac{d^2 \bfLp}{\dlw} L^{\prime}_c L^{\prime}_f  \int \frac{d^2 \bfLpp}{\dlw} L^{\prime \prime}_g L^{\prime \prime}_b  \int \frac{d^2 \bfLppp}{\dlw} L^{\prime \prime \prime}_f L^{\prime \prime \prime}_d \, \delta_D (\bfL + \bfLp - \bfell) \delta_D (\bfellp + \bfLp + \bfLppp)  \\
\times \frac{1}{\chi^3 \chipp \chip^3 \chippp} \left\langle {\Psi} (\chi,\bfL) {\Psi} (\chip,\bfLp) {\Psi} (\chipp,\bfLpp) {\Psi} (\chippp,\bfLppp) \right\rangle + \lbrace ab \leftrightarrow cd \rbrace .
\end{multline}
\n
As before, upon calculating the Wick contractions we find a number of convenient cancellations. As the structure of the Wick contractions is identical to the 22D correction, we will note reproduce the vanishing contributions. The non-vanishing Wick contraction is given by
\begin{align}
\langle
\contraction{}{{\Psi}}{{}_1(\mathbf{q}_1)\>\>\>\>{}}{{\Psi}}
  \nomathglue{%
     {\Psi} (\chi,\bfL)\>\>{}{\Psi} (\chip, \bfLp)}
\bcontraction{}{{\Psi}}{{}(\chipp,\bfLpp)\>{}}{{\Psi}} \;
  \nomathglue{%
     {\Psi}(\chipp , \bfLpp)\>{}{\Psi} (\chippp , \bfLppp)}
\rangle
\propto \delta_D (\chi - \chip) \, \delta_D (\chipp - \chippp) \, \delta_D (\bfL + \bfLp) \, \delta_D (\bfLpp + \bfLppp) ,
\end{align}
\n
which produces a term $W^2 (\chi,\chi_s) W^2 (\chipp,\chi)$. This final contribution leads to $\bfLp \rightarrow -\bfL$ and $\bfLppp \rightarrow -\bfLpp$, as before. This time, however, we find a kernel $\propto \bfL^2 \, \bfL^{\prime \prime 2} \, ( \bfL \cdot \bfLpp)^2$. When coupled with the delta functions on the second line of Eq.~(\ref{eqn:22LFull}), this kernel reduces to a term $\propto \bfL^2 \left( \bfell - \bfL \right)^2 \left[ \bfL \cdot (\bfell - \bfL) \right]^2$ where the integration is again carried out over $\bfL$.

The 22L corrections can therefore be shown to reduce to the following contributions \cite{Cooray:2002mj,Krause:2009yr}
\begin{align}
\Delta C^{22L}_{abcd} (\bfell) &= 16 \int \frac{d^2 \bfellp}{(2 \pi)^2} \, \ell_a^{\prime} \ell_b^{\prime \prime} \ell_c^{\prime} \ell_d^{\prime \prime} \left( \bfellp \cdot \bfellpp \right)^2 \, M (\ell^{\prime} , \ell^{\prime \prime}) ,
\end{align}
\n
which simplifies to an integral over a common kernel weighted by the appropriate geometrical factors
\begin{align}
\Delta C^{\mu \nu, 22L}_{\ell}
&= 4 \int \frac{d^2 \bfellp}{(2 \pi)^2} \, \ell^{\prime 2} \, \ell^{\prime \prime 2} G^{(22L)}_{\mu} G^{(22L)}_{\nu} \left( \, \bfellp \cdot \bfellpp \, \right)^2 M(\ell^{\prime} , \ell^{\prime \prime}) ,
\end{align}
\n
where the 22L geometrical factors are defined to be \cite{Cooray:2002mj}
\begin{align}
G^{(22L)}_{\kappa} = \cos (\varphi_{\ell^{\prime}} - \varphi_{\ell^{\prime \prime}}), \quad G^{(22L)}_{\gamma_E} = \cos (\varphi_{\ell^{\prime}} + \varphi_{\ell^{\prime \prime}}), \quad G^{(22L)}_{\gamma_B} = \sin (\varphi_{\ell^{\prime}} + \varphi_{\ell^{\prime \prime}}) , \quad G^{(22L)}_{\omega} = \sin (\varphi_{\ell^{\prime}} - \varphi_{\ell^{\prime \prime}}) .
\end{align}
\n
The 22 lens-lens coupling induces power in all weak lensing observables but the parity violating terms will still oscillate causing non-trivial cancellations after integration over the azimuthal angles.

\subsubsection{22X Correction}
The final second-second order correction arises from considering the cross terms between corrections due to relaxing the Born approximation as well as lens-lens coupling terms. The derivation of these terms proceeds in exactly the same way as discussed above. The kernel for this integration has the schematic form
\begin{multline}
\label{eqn:22XFull}
\langle \psi^{(22D)}_{ab}  (\bfell,\chi_s) \, \psi^{(22L)}_{cd} (\bfellp , \chi_s) \rangle = \int^{\chi_s}_0 d\chi \, \int^{\chi_s}_0 d \chip \, \int^{\chip}_0 d \chipp \int^{\chip}_0 d \chippp \, \chi^3 W(\chi,\chi_s) \chi^{\prime 2} W(\chip,\chi_s) \chipp W(\chipp,\chip) \chi^{\prime \prime 3} W(\chippp,\chip) \\
 \times (2 \pi)^4 \int \frac{d^2 \bfL}{\dlw} L_a L_b L_f  \int \frac{d^2 \bfLp}{\dlw} L^{\prime}_c L^{\prime}_g  \int \frac{d^2 \bfLpp}{\dlw} L^{\prime \prime}_f  \int \frac{d^2 \bfLppp}{\dlw} L^{\prime \prime \prime}_g L^{\prime \prime \prime}_d \,\delta_D (\bfL + \bfLp - \bfell) \delta_D (\bfellp + \bfLp + \bfLppp) \\
 \times \frac{1}{\chi^3 \chipp \chip^2 \chippp^2} \left\langle {\Psi} (\chi,\bfL) {\Psi} (\chip,\bfLp) {\Psi} (\chipp,\bfLpp) {\Psi} (\chippp,\bfLppp) \right\rangle + \lbrace ab \leftrightarrow cd \rbrace ,
\end{multline}
\n
which will lead to only one surviving Wick contraction. This non-vanishing term is again
\begin{align}
\langle
\contraction{}{{\Psi}}{{}_1(\mathbf{q}_1)\>\>\>\>{}}{{\Psi}}
  \nomathglue{%
     {\Psi} (\chi,\bfL)\>\>{}{\Psi} (\chip, \bfLp)}
\bcontraction{}{{\Psi}}{{}(\chipp,\bfLpp)\>{}}{{\Psi}} \;
  \nomathglue{%
     {\Psi}(\chipp , \bfLpp)\>{}{\Psi} (\chippp , \bfLppp)}
\rangle
\propto \delta_D (\chi - \chip) \, \delta_D (\chipp - \chippp) \, \delta_D (\bfL + \bfLp) \, \delta_D (\bfLpp + \bfLppp) .
\end{align}
\n
The 22X corrections are therefore given by \cite{Cooray:2002mj,Krause:2009yr}
\begin{align}
\Delta C^{22X}_{abcd} (\ell) &= 16 \int \frac{d^2 \bfellp}{(2 \pi)^2} \, \left( \ell_a^{\prime} \ell_b^{\prime} \ell_c^{\prime} \ell_d^{\prime \prime} + \ell_a^{\prime} \ell_b^{\prime \prime} \ell_c^{\prime} \ell_d^{\prime} \right) \left( \bfellp \cdot \bfellpp \right)^2 \, M (\ell^{\prime} , \ell^{\prime \prime}) ,
\end{align}
\n
which can be re-expressed as
\begin{align}
\Delta C^{\mu\nu, 22X}_{\ell} &= 4 \int \frac{d^2 \bfellp}{(2 \pi)^2} \, \ell^{\prime 3} \, \ell^{\prime \prime} G^{(22X)}_{\mu} G^{(22X)}_{\nu} \left[ \, \bfellp \cdot \bfellpp \, \right]^2 M(\ell^{\prime} , \ell^{\prime \prime}) ,
\end{align}
\n
where the geometric factors are given by \cite{Cooray:2002mj}
\begin{align}
G^{(22X)}_{\mu} G^{(22X)}_{\nu} &= G^{(22D)}_{\mu} G^{(22L)}_{\nu} + G^{(22L)}_{\mu} G^{(22D)}_{\nu} .
\end{align}

\subsubsection{22 Corrections}
The corrections from all the contributions of second-second order can be re-written in a more compact form by noting that a number of convenient cancellations occurs. By summing over the deflection, lens-lens and deflection-lens coupling terms, the total 22 corrections will reduce to \cite{Krause:2009yr}
\begin{align}
\Delta C^{\mu \nu, 22}_{\ell} &= 4 \int \frac{d^2 \bfellp}{(2 \pi)^2} \, \ell^{2} \, \ell^{\prime 2} \; G^{(22)}_{\mu} G^{(22)}_{\nu} \left[ \, \bfellp \cdot \bfellpp \, \right]^2 M(\ell^{\prime} , \ell^{\prime \prime}) ,
\end{align}
\n
with the 22 geometric factors now being given by
\begin{align}
G^{(22)}_{\kappa} = \cos (\varphi_{\ell^{\prime}}), \quad G^{(22)}_{\gamma_E} = \cos(\varphi_{\ell^{\prime}}), \quad G^{(22)}_{\gamma_B} = \sin (\varphi_{\ell^{\prime}}) , \quad G^{(22)}_{\omega} = \sin (\varphi_{\ell^{\prime}}) .
\end{align}
\n
This agrees with the expected equivalence between the E-mode shear and convergence as well as an equivalence between the B-mode shear and curl power spectra \cite{Krause:2009yr}.

\subsubsection{13D Correction}
Under the Limber approximation, we can show that this is the only non-vanishing 13 correction that will survive \cite{Krause:2009yr}. The 13D term encapsulates the ray-deflection corrections to the geodesic path of the photon bundle. The first-third order term can be shown to reduce to \cite{Cooray:2002mj,Krause:2009yr}
\begin{align}
\Delta C^{\mu \nu, 13D}_{\ell} &= -4 \int \frac{d^2 \bfellp}{(2 \pi)^2} \, \ell^{4} G^{(13D)}_{\mu} G^{(13D)}_{\nu} \left[ \, \bfell \cdot \bfellp \, \right]^2 M(\ell , \ell^{\prime}) ,
\end{align}
\n
where we have introduced the geometric factors \cite{Cooray:2002mj}
\begin{align}
G^{(13D)}_{\kappa} = 1, \quad G^{(13D)}_{\gamma_E} = 1, \quad G^{(13D)}_{\gamma_B} = 0 , \quad G^{(13D)}_{\omega} = 0.
\end{align}

\subsubsection{13X Correction}
The contributions from the 13X correction can be shown to vanish under the Limber approximation. This can be seen by looking at the structure of the power spectrum for the 13X terms
\begin{multline}
\label{eqn:13XFull}
\langle \psi^{(1)}_{ab}  (\bfell,\chi_s) \, \psi^{(3X)}_{cd} (\bfellp , \chi_s) \rangle = \int^{\chi_s}_0 d\chi \, \int^{\chi_s}_0 d \chip \, \int^{\chip}_0 d \chipp \int^{\chip}_0 d \chippp \, \chi^2 W(\chi,\chi_s) \chi^{\prime 3} W(\chip,\chi_s) \chi^{\prime \prime 2} W(\chipp,\chip) \chipp W(\chippp,\chip) \\
\times (2 \pi)^4 \int \frac{d^2 \bfL}{\dlw} L_a L_b  \int \frac{d^2 \bfLp}{\dlw} L^{\prime}_a L^{\prime}_c L^{\prime}_d \int \frac{d^2 \bfLpp}{\dlw} L^{\prime \prime}_b L^{\prime \prime}_c \int \frac{d^2 \bfLppp}{\dlw} L^{\prime \prime \prime}_d \, \delta_D (\bfL - \bfell) \delta_D (\bfellp + \bfLp + \bfLpp + \bfLppp)   \\ \times
\frac{1}{\chi^2 \chip^3 \chipp^2 \chippp} \left\langle {\Psi} (\chi,\bfL) {\Psi} (\chip,\bfLp) {\Psi} (\chipp,\bfLpp) {\Psi} (\chippp,\bfLppp) \right\rangle + \lbrace ab \leftrightarrow cd \rbrace .
\end{multline}
\n
Evaluating the Wick contraction on the angular Fourier transform of the potentials ${\Psi}$ leads to two vanishing and one non-vanishing contribution. The first vanishing Wick contraction is given by
\begin{align}
\langle
\contraction{}{{\Psi}}{{}_1(\mathbf{q}_1)\>\>\>\>{}}{{\Psi}}
  \nomathglue{%
     {\Psi} (\chi,\bfL)\>\>{}{\Psi} (\chipp, \bfLpp)}
\bcontraction{}{{\Psi}}{{}(\chipp,\bfLpp)\>{}}{{\Psi}} \;
  \nomathglue{%
     {\Psi}(\chip , \bfLp)\>\>{}{\Psi} (\chippp , \bfLppp)}
\rangle
\propto \delta_D (\chi - \chipp) \, \delta_D (\chip - \chippp) \, \delta_D (\bfL + \bfLppp) \, \delta_D (\bfLp + \bfLpp) ,
\end{align}
\n
as the delta functions imply that we have the following term in the integral $\delta_D (\chip - \chippp) W(\chippp,\chip) = 0$, hence this vanishes. The second vanishing Wick contraction is given by
\begin{align}
\langle
\contraction{}{{\Psi}}{{}_1(\mathbf{q}_1)\>\>\>\>{}}{{\Psi}}
  \nomathglue{%
     {\Psi} (\chi,\bfL)\>\>{}{\Psi} (\chippp, \bfLppp)}
\bcontraction{}{{\Psi}}{{}(\chipp,\bfLpp)\>{}}{{\Psi}} \;
  \nomathglue{%
     {\Psi}(\chip , \bfLp)\>\>{}{\Psi} (\chipp , \bfLpp)}
\rangle
\propto \delta_D (\chi - \chippp) \, \delta_D (\chip - \chipp) \, \delta_D (\bfL + \bfLpp) \, \delta_D (\bfLp + \bfLppp) ,
\end{align}
\n
which again implies that we have a term of the form $\delta_D (\chip - \chipp) W(\chipp,\chip) = 0$, hence this also vanishes. The non-vanishing Wick contribution is given by
\begin{align}
\langle
\contraction{}{{\Psi}}{{}_1(\mathbf{q}_1)\>\>\>\>{}}{{\Psi}}
  \nomathglue{%
     {\Psi} (\chi,\bfL)\>\>{}{\Psi} (\chip, \bfLp)}
\bcontraction{}{{\Psi}}{{}(\chipp,\bfLpp)\>{}}{{\Psi}} \;
  \nomathglue{%
     {\Psi}(\chipp , \bfLpp)\>{}{\Psi} (\chippp , \bfLppp)}
\rangle
\propto \delta_D (\chi - \chip) \, \delta_D (\chipp - \chippp) \, \delta_D (\bfL + \bfLp) \, \delta_D (\bfLpp + \bfLppp) .
\end{align}
\n
This can be shown to collapse to a contribution of the form
\begin{multline}
\langle \psi^{(1)}_{ab}  (\bfell,\chi_s) \, \psi^{(3X)}_{cd} (\bfellp , \chi_s) \rangle = \delta_D (\bfell + \bfellp) \int^{\chi_s}_0 d\chi \, \int^{\chip}_0 d \chipp \int^{\chi}_0 d \chippp \, \frac{W^2(\chi,\chi_s)}{\chi^2} \frac{W^2(\chipp,\chi)}{\chipp^2}  \\
 \times \int \frac{d^2 \bfLpp}{\dlw} \ell_a \ell_b \ell_a \ell_c \ell_d L^{\prime \prime}_b L^{\prime \prime}_c L^{\prime \prime}_d \; P_{\Psi \Psi} \left( \frac{L}{\chi} ; \chi \right) \; P_{\Psi \Psi} \left( \frac{L^{\prime}}{\chip} ; \chip \right) + \lbrace ab \leftrightarrow cd \rbrace .
\end{multline}
\n
This contribution is odd parity under $\bfLpp \rightarrow -\bfLpp$ and will therefore vanish \cite{Cooray:2002mj,Krause:2009yr}.

\subsubsection{13L, 13X1 and 13X2 Corrections}
The 13L term also vanishes due to the implicit restriction $\chipp < \chip < \chi$ in the integrals \cite{Krause:2009yr}. If we consider the structure of the 13L correction we see that
\begin{multline}
\langle \psi^{(1)}_{ab}  (\bfell,\chi_s) \, \psi^{(3D)}_{cd} (\bfellp , \chi_s) \rangle = \int^{\chi_s}_0 d \chi \, \int^{\chi_s} d \chip \, \int^{\chip}_0 d \chipp \int^{\chipp} d \chippp \, \chi^2 W(\chi,\chi_s) \chi^{\prime 2} W(\chip , \chi_s) \chi^{\prime\prime 2} W(\chipp,\chip) \chi^{\prime\prime\prime 2} W(\chippp,\chipp) \\
\nonumber \times (2\pi)^4  \int \frac{d^2 \bfL}{\dlw} L_a L_b \int \frac{d^2 \bfLp}{\dlw} L^{\prime}_c L^{\prime}_f \int \frac{d^2 \bfLpp}{\dlw} L^{\prime \prime}_f L^{\prime \prime}_g \, \int \frac{d^2 \bfLppp}{\dlw} L^{\prime \prime \prime}_g L^{\prime \prime \prime}_c \, \delta_D (\bfL - \bfell) \delta_D (\bfellp + \bfLp + \bfLpp + \bfLppp)  \\
\nonumber  \times \frac{1}{\chi^2 \chip^2 \chipp^2 \chippp^2}  \; \langle {\Psi} (\chi , \bfL) {\Psi} (\chip , \bfLp) {\Psi} (\chipp, \bfLpp) {\Psi} (\chippp, \bfLppp) \rangle + \lbrace{ab} \leftrightarrow {cd} \rbrace .
\end{multline}
\n
The Wick contractions of the angular Fourier transform of the potentials leads to terms such as
\begin{align}
\langle
\contraction{}{{\Psi}}{{}_1(\mathbf{q}_1)\>\>\>\>{}}{{\Psi}}
  \nomathglue{%
     {\Psi} (\chi,\bfL)\>\>{}{\Psi} (\chip, \bfLp)}
\bcontraction{}{{\Psi}}{{}(\chipp,\bfLpp)\>\>\>{}}{{\Psi}} \;
  \nomathglue{%
     {\Psi}(\chipp , \bfLpp)\>\>{}{\Psi} (\chippp , \bfLppp)}
\rangle
\propto \delta_D (\chi - \chip) \, \delta_D (\chipp - \chippp) \, \delta_D (\bfL + \bfLp) \, \delta_D (\bfLpp + \bfLppp)  .
\end{align}
\n
The delta function $\delta_D (\chipp - \chippp)$ will cause the integral to vanish as $W(\chi,\chi) = 0$. The second Wick terms lead to
\begin{align}
\langle
\contraction{}{{\Psi}}{{}_1(\mathbf{q}_1)\>\>\>\>{}}{{\Psi}}
  \nomathglue{%
     {\Psi} (\chi,\bfL)\>\>{}{\Psi} (\chipp, \bfLpp)}
\bcontraction{}{{\Psi}}{{}(\chipp,\bfLpp)\>{}}{{\Psi}} \;
  \nomathglue{%
     {\Psi}(\chip , \bfLp)\>\>{}{\Psi} (\chippp , \bfLppp)}
\rangle
\propto \delta_D (\chi - \chipp) \, \delta_D (\chip - \chippp) \, \delta_D (\bfL + \bfLpp) \, \delta_D (\bfLp + \bfLppp) ,
\end{align}
\n
which also kills the integral as $W(\chipp,\chip) W(\chippp,\chipp) \delta_D (\chip-\chippp) \delta_D (\chi - \chipp) = W(\chi,\chip) W(\chip,\chi) = 0$. The final Wick term leads to
\begin{align}
\langle
\contraction{}{{\Psi}}{{}_1(\mathbf{q}_1)\>\>\>\>{}}{{\Psi}}
  \nomathglue{%
     {\Psi} (\chi,\bfL)\>\>{}{\Psi} (\chippp, \bfLppp)}
\bcontraction{}{{\Psi}}{{}(\chipp,\bfLpp)\>{}}{{\Psi}} \;
  \nomathglue{%
     {\Psi}(\chip , \bfLp)\>\>{}{\Psi} (\chipp , \bfLpp)}
\rangle
\propto \delta_D (\chi - \chippp) \, \delta_D (\chip - \chipp) \, \delta_D (\bfL + \bfLppp) \, \delta_D (\bfLp + \bfLpp) ,
\end{align}
\n
for which $W(\chipp,\chip) W(\chippp,\chipp) \delta_D (\chi-\chippp) \delta_D (\chip - \chipp) = W(\chip,\chip) W(\chi,\chip) = 0$ as $W(\chi,\chi) = 0$. Similarly, the 13X1 and 13X2 corrections lead to Wick contractions that kill off the integral and consequentially none of these terms contribute to the post-Born corrections at $\mathcal{O}(\Psi^4)$ under the Limber approximation. This result is in agreement with \cite{Krause:2009yr} who also argue that the $\psi^{(3C)}_{ab}$ term, in their notation, necessarily vanishes due to the restriction $\chipp < \chip < \chi$ in the integrals.

\subsection{Power spectra}

\label{appendix:power}

Now that we have the corrections to the deformation tensor to the appropriate order, we can start to construct the power spectra up to $\mathcal{O}(\Psi^4)$. Statistical homogeneity demands that the two-point correlation of the deformation tensor obeys
\begin{align}
\left\langle \psi_{ab}^{\ast} (\bfell) \, \psi_{cd} (\bfellp) \right\rangle &= \left( 2 \pi^2 \right) \, \delta_D \left( \bfell - \bfellp \right) \, C_{abcd} ( \bfell ) .
\end{align}
\n
This object can then be used to define the two-point statistics of the various lensing observables by the appropriate contractions. The power and cross-spectra of the observable fields $\alpha , \beta = \kappa, \gamma_E, \gamma_B$ and $\omega$ are defined by \cite{Cooray:2002mj}
\begin{align}
\calC_{\ell}^{\kappa \kappa} &= \frac{1}{4} \left[ C_{1111} + 2 C_{1122} + C_{2222} \right] ,\\
\calC_{\ell}^{\gamma_E \gamma_E} &= \frac{1}{4} \left[ C_{1111} - 2 C_{1122} + C_{2222} \right] , \\
\calC_{\ell}^{\gamma_B \gamma_B} &= \frac{1}{4} \left[ C_{1212} + 2 C_{1221} + C_{2121} \right]  ,\\
\calC_{\ell}^{\omega \omega} &= \frac{1}{4} \left[ C_{1212} - 2 C_{1221} + C_{2121} \right] .
\end{align}
\n
For CMB weak lensing, we will only require the power spectrum for the lensing potential $\calC^{\phi \phi}_{\ell}$ and the curl potential $\calC^{\Omega \Omega}_{\ell}$. Under the Limber approximation, the power spectra for the lensing potentials to order $\mathcal{O}(\Psi^4)$ simplifies considerably. The only surviving 13 corrections correspond to the ray-deflection terms \cite{Cooray:2002mj,Krause:2009yr}. This term is just given by
\begin{align}
\Delta C^{\mu \nu, 13D}_{\ell} &= -4 \int \frac{d^2 \bfellp}{(2 \pi)^2} \, \ell^{4} G^{(13D)}_{\mu} G^{(13D)}_{\nu} \left[ \, \bfell \cdot \bfellp \, \right]^2 M(\ell , \ell^{\prime}) ,
\end{align}
\n
where we have introduced the geometric factors \cite{Cooray:2002mj}
\begin{align}
G^{(13D)}_{\kappa} = 1, \quad G^{(13D)}_{\gamma_E} = 1, \quad G^{(13D)}_{\gamma_B} = 0 , \quad G^{(13D)}_{\omega} = 0,
\end{align}
\n
and we have adopted the standard notation in the literature \cite{Cooray:2002mj,Krause:2009yr} and introduced a mode-coupling kernel
\begin{align}
\label{eqn:MKH}
M (\ell , \ell^{\prime}) &= \int^{\chi_s}_0 d \chi \, \frac{W^2 (\chi,\chi_s)}{\chi^2} \int^{\chi}_0 d \chi^{\prime} \, \frac{W^2 (\chi^{\prime} , \chi)}{\chi^{\prime 2}} \, P_{\Psi \Psi} \left( \frac{\ell^{\prime}}{\chi^{\prime}} ; \chi^{\prime} \right) \,  P_{\Psi \Psi} \left( \frac{\ell}{\chi} ; \chi \right) .
\end{align}

For the 22 corrections, we sum over the ray-deflection, lens-lens and deflection-lens coupling terms to find the total effective correction. This is given by \cite{Cooray:2002mj,Krause:2009yr}
\begin{align}
\Delta C^{\alpha \beta, 22}_{\ell} &= 4 \int \frac{d^2 \bfellp}{(2 \pi)^2} \, \ell^{2} \, \ell^{\prime 2} \; G^{(22)}_{\alpha} G^{(22)}_{\gamma_B} \left[ \, \bfellp \cdot \bfellpp \, \right]^2 M(\ell^{\prime} , \ell^{\prime \prime}) ,
\end{align}
\n
where the 22 geometric factors are given by
\begin{align}
G^{(22)}_{\kappa} = \cos (\varphi_{\ell^{\prime}}), \quad G^{(22)}_{\gamma_E} = \cos(\varphi_{\ell^{\prime}}), \quad G^{(22)}_{\gamma_B} = \sin (\varphi_{\ell^{\prime}}) , \quad G^{(22)}_{\omega} = \sin (\varphi_{\ell^{\prime}}) .
\end{align}
\n
As can be seen above, there is an equivalence between the E-mode shear and the convergence as well as an equivalence between the B-mode shear and curl power spectra due to the fact that the deflection angle only has two underlying degrees of freedom. Note that whilst we have opted to use the Limber approximation, a full-sky treatment of these terms is possible but would be more involved \cite{Bernardeau:2009bm}. The role of full-sky corrections in CMB weak lensing is left to future work.

%%%%%%%%%%%%%%%%%%
\allowdisplaybreaks
%%%%%%%%%%%%%%%%%%

%%%%%%%%%%%%%%%%%%%%%%%% References %%%%%%%%%%%%%%%%%%%%%%%%%

\bibliography{PostBorn,antony,cosmomc}

\end{document}